\documentclass[reqno]{amsart}

\usepackage{amsmath}
\usepackage{amsfonts}
\usepackage{amsthm}

\numberwithin{equation}{section}

\newcommand{\supp}{\operatorname{supp}}
\newcommand{\Spec}{\operatorname{Spec}}

\newcommand{\Div}{{\operatorname{div}\,}}

\newcommand{\dist}{{\operatorname{dist}}}

\newcommand{\curl}{{\operatorname{curl}\,}}

\newcommand{\Hess}{{\operatorname{Hess}\,}}

\newtheorem{theorem}{Theorem}[section]
\newtheorem{thm}[theorem]{Theorem}
\newtheorem{lemma}[theorem]{Lemma}
\newtheorem{proposition}[theorem]{Proposition}
\newtheorem{prop}[theorem]{Proposition}

\newtheorem{remark}[theorem]{Remark}
\newtheorem{corollary}[theorem]{Corollary}

\newtheorem{cor}[theorem]{Corollary}

\newtheorem{assumption}[theorem]{Assumption}

\title{On the third critical field in Ginzburg-Landau theory}

\author{S. Fournais}
\author{B. Helffer}

\thanks{The two authors are supported by the European Research Network
`Postdoctoral Training Program in Mathematical Analysis of Large
Quantum Systems' with contract number HPRN-CT-2002-00277, and the ESF
Scientific Programme in Spectral Theory and Partial Differential
Equations (SPECT). Part of this work was carried out while S.F. visited CIMAT, Mexico} 
\address[S. Fournais]{CNRS and Laboratoire de
Math\'{e}matiques UMR CNRS 8628\\ Universit\'{e} Paris-Sud - B\^{a}t 425\\
F-91405 Orsay Cedex\\ France.
}
\email{soeren.fournais@math.u-psud.fr}

\address[B. Helffer]{Laboratoire de
Math\'{e}matiques UMR CNRS 8628\\ Universit\'{e} Paris-Sud - B\^{a}t 425\\
F-91405 Orsay Cedex\\ France.
}
\email{bernard.helffer@math.u-psud.fr}

\date{\today}

\begin{document}

\bibliographystyle{plain}

\begin{abstract}
Using recent results by the authors on the spectral asymptotics of the Neumann Laplacian with magnetic field, we give precise estimates on the critical field, $H_{C_3}$, describing the appearance of superconductivity in superconductors of type II. 
Furthermore, we prove that the local and global definitions of this field coincide.
Near $H_{C_3}$ only a small part, near the boundary points where the curvature is maximal, of the sample carries superconductivity. We give precise estimates on the size of this zone and decay estimates in both the normal (to the boundary) and parallel variables.
\end{abstract} 

\maketitle
\tableofcontents

\section{Introduction}
~
\subsection{Setup and results for general domains}~\\
Our main motivation comes from superconductivity. As appeared 
 from  the works of Bernoff-Sternberg~\cite{BeSt}, Lu-Pan~\cite{LuPa1, LuPa2, LuPa3, LuPa5}, and Helffer-Pan~\cite{He-Pan}, the determination of 
the lowest eigenvalues of the magnetic Schr\"{o}dinger operator is
crucial for a detailed description of the nucleation of
superconductivity (on the boundary) for superconductors of Type II and
for accurate estimates of the critical field $H_{C_3}$. If the determination
of the complete asymptotics of the lowest eigenvalues of the Schr\"odinger operators
was essentially achieved (except for exponentially small effects) in the two-dimensional case 
with the works of \cite{He-Mo} and \cite{FournaisHelffer2}, what remained to be determined
was the corresponding asymptotics for the critical field. We will actually
obtain much more and clarify the links between the various definitions
of critical fields considered in the mathematical or physical literature
and supposed to define the right critical field.\\

The Ginzburg-Landau functional is given by
\begin{multline}
\label{eq:GL_F}
{\mathcal E}[\psi,\vec{A}] = {\mathcal
E}_{\kappa,H}[\psi,\vec{A}]  =
\int_{\Omega} \Big\{ |p_{\kappa H \vec{A}}\psi|^2 
- \kappa^2|\psi|^2
+\frac{\kappa^2}{2}|\psi|^4 \\
+ \kappa^2 H^2
|\curl \vec{A} - 1|^2\Big\}\,dx\;,
\end{multline}
with
$ (\psi, \vec{A})  \in W^{1,2}(\Omega;{\mathbb C})\times
W^{1,2}(\Omega;{\mathbb R}^2)$ and where $p_{\vec{A}} = (-i\nabla- \vec{A})$\,. 

We fix the choice of gauge by imposing that 
\begin{align}
\label{eq:gauge}
\Div \vec{A} &= 0 \quad \text{ in } \Omega\;, & \vec{A} \cdot \nu = 0 \quad \text{ on } \partial \Omega\;.
\end{align}
By variation around a minimum for ${\mathcal E}_{\kappa,H}$ we find that minimizers $(\psi, \vec{A})$ satisfy the Ginzburg-Landau equations,
\begin{subequations}
\label{eq:GL}
\begin{align}
\left.\begin{array}{c}
p_{\kappa H \vec{A}}^2\psi =
\kappa^2(1-|\psi|^2)\psi \\
\label{eq:equationA}
\curl^2 \vec{A} =-\tfrac{i}{2\kappa H}(\overline{\psi} \nabla
\psi - \psi \nabla \overline{\psi}) - |\psi|^2 \vec{A}
\end{array}\right\} &\quad \text{ in } \quad \Omega \, ;\\
\left. \begin{array}{c}
(p_{\kappa H \vec{A}} \psi) \cdot \nu = 0 \\
\curl \vec{A} - 1 = 0
\end{array} \right\} &\quad \text{ on } \quad \partial\Omega \, .
\end{align}
\end{subequations}
Here $\curl (A_1,A_2) = \partial_{x_1}A_2 -
\partial_{x_2}A_1$, and
$$
\curl^2 {\vec A} =
(\partial_{x_2}(\curl {\vec A}),-\partial_{x_1}(\curl {\vec
A})) \, .
$$

Let $\vec{F}$ denote the vector potential generating the constant exterior magnetic field
\begin{align}
\label{eq:F}
&\left.
\begin{array}{c}
\Div \vec{F} = 0\quad\\
\curl \vec{F} =1\quad
\end{array} \right\}
\quad \text{ in } \Omega\;, & \vec{F} \cdot \nu = 0 \quad \text{ on } \partial \Omega\;.
\end{align}

It is known that, for given values of the the parameters $\kappa, H$, the functional ${\mathcal E}$ has (possibly non-unique) minimizers. However, after some analysis of the functional, one finds (see \cite{Giorgi-Phillips} for details) that given $\kappa$ there exists $H(\kappa)$ such that if $H>H(\kappa)$ then $(0,\vec{F})$ is the only minimizer of ${\mathcal E}_{\kappa,H}$ (up to change of gauge).

Following Lu and Pan \cite{LuPa1}, we can therefore define
\begin{align}
H_{C_3}(\kappa) = \inf\{ H>0 \;:\; (0, \vec{F}) \text{ is a minimizer of } {\mathcal E}_{\kappa,H}\}\;.
\end{align}
In the physical interpretation of a minimizer $(\psi,\vec{A})$, $|\psi(x)|$ is a measure of the superconducting properties of the material near the point $x$. Therefore, $H_{C_3}(\kappa)$ is the value of the external magnetic field, $H$, at which the material loses its superconductivity completely.

A central question in the mathematical treatment of Type II\footnote{Superconductors of Type II are the ones for which $\kappa$ (in our units) is large.} superconductors is to establish the asymptotic behavior of $H_{C_3}(\kappa)$ for large $\kappa$. We will also be concerned with this and will, in particular, describe how $H_{C_3}(\kappa)$ can be determined by the study of a linear problem.

Our first result is the following strengthening of a result in \cite{He-Pan}.

\begin{thm}
\label{thm:GeneralHG3}~\\
Suppose $\Omega$ is a bounded simply-connected domain in ${\mathbb R}^2$ with smooth boundary. 
Let $k_{{\rm max}}$ be the maximal curvature of $\partial \Omega$.
Then
\begin{align}
\label{eq:GenAsymp}
  H_{C_3}(\kappa) = \frac{\kappa}{\Theta_0} + \frac{C_1}{\Theta_0^{\frac{3}{2}}} k_{{\rm max}} + {\mathcal O}(\kappa^{-\frac{1}{2}}) \;,
\end{align}
where $C_1, \Theta_0$ are universal constants.

When $\Omega$ is a disc we get the improved estimate
\begin{align}
\label{eq:DiscAsymp}
  H_{C_3}(\kappa) = \frac{\kappa}{\Theta_0} + \frac{C_1}{\Theta_0^{\frac{3}{2}}} k_{{\rm max}} + {\mathcal O}(\kappa^{-1}) \;.
\end{align}
\end{thm}

\begin{remark}~\\
The constants $\Theta_0, C_1$ are defined in terms of auxiliary spectral problems which are presented in Appendix~\ref{AppA}.
\end{remark}

The proof of Theorem~\ref{thm:GeneralHG3} is given in Section~\ref{sec:GenHC3} below.

$\,$
\begin{remark}~\\
The improvement in \eqref{eq:GenAsymp} compared to \cite[Theorem~1.1]{He-Pan} is in the estimate on the remainder which are ${\mathcal O}(\kappa^{-\frac{1}{2}})$ instead of 
${\mathcal O}(\kappa^{-\frac{1}{3}})$. Our result is optimal in the sense that the next term depends on detailed geometric properties of the boundary. 
We believe that (at least `generically') the next term in $H_{C_3}(\kappa)$ is of the form $c_0 \kappa^{-a}$ where both $c_0\in {\mathbb R}$ and $a\geq \frac{1}{2}$ depend on $\partial \Omega$.
In order to expand $H_{C_3}$ to higher orders we will impose a geometric condition, Assumption~\ref{assump:GenNondegen} below, on $\Omega$.
\end{remark}

Our second result is a precise estimate on the size of the superconducting region in the case where $H$ is close to, but below, $H_{C_3}$. To state it we need a bit of notation concerning the boundary, $\partial \Omega$. Let $\gamma : {\mathbb R}/|\partial \Omega| \rightarrow {\mathbb R}^2$ be a (counter-clockwise) parametrization of $\partial \Omega$ with $|\gamma'(s)| = 1$. For $s \in {\mathbb R}/|\partial \Omega|$ we denote by $k(s)$ the curvature of $\partial \Omega$ at the point $\gamma(s)$. 
For more discussion of these boundary coordinates, see Appendix~\ref{bdry}.
Furthermore, we introduce
\begin{align}
\label{eq:Ks}
k_{{\rm max}}&:= \max_{s \in {\mathbb R}/|\partial \Omega|} k(s)\;, &
K(s) &:= k_{{\rm max}} - k(s)\;.
\end{align}
Furthermore, we define the coordinate $t=t(x)$ that measures the distance to the boundary
$$
t(x) := \dist(x, \partial \Omega)\;.
$$
Let $\nu(s)$ be the interior normal vector to $\partial \Omega$ at the point $\gamma(s)$ and define
$\Phi : {\mathbb R}/|\partial \Omega| \times (0,t_0) \rightarrow \Omega$ by
$$
\Phi(s,t) = \gamma(s) + t \nu(s)\;.
$$
Then, for $t_0$ sufficiently small, $\Phi$ is a diffeomorphism with image
$$
\Phi\big( {\mathbb R}/|\partial \Omega| \times (0,t_0) \big) = 
\{ x \in \Omega \big | \dist(x, \partial \Omega) < t_0 \} \;.
$$
Furthermore, $t(\Phi(s,t)) = t$.
Thus, in a neighborhood of the boundary, the function $s=s(x)$ is defined (by $(s(x), t(x)) = \Phi^{-1}(x)$). 

From the work of Helffer-Morame \cite{He-Mo} (see also Helffer-Pan \cite{He-Pan} for the non-linear case) we know that minimizers of the Ginzburg-Landau functional are exponentially localized to a region near the boundary (see Theorem~\ref{thm:agmon} for a restatement of their results). Here we prove that minimizers are also localized in the tangential variable to a small zone around the points of maximum curvature. The size of that zone depends on the order to which the derivatives of the curvature vanishes at such points.
Our estimate is an improvement of a similar estimate in \cite{He-Pan}.

\begin{thm}[Tangential Agmon estimates (non-linear case)]
\label{thm:nucleation}~\\
Let $\Omega$ be a bounded simply-connected domain in ${\mathbb R}^2$ with smooth boundary. 
Let $(\psi, \vec{A}) = (\psi_{\kappa, H}, \vec{A}_{\kappa, H})$ be a family of minimizers of the Ginzburg-Landau functional depending on the parameters $\kappa, H$.
We suppose that $H = H(\kappa)$ in such a way that
$\rho := H_{C_3}(\kappa) - H$ satisfies $0 < \rho = o(1)$ as $\kappa \rightarrow \infty$.
Then there exist $\alpha, C >0$ such that if $\kappa > C$, then
\begin{align}
\int_{\Omega} \chi_1^2( \kappa^{\frac{1}{4}} t) e^{2 \alpha \sqrt{\kappa} K(s) }|\psi(x)|^2 \,dx \leq C e^{C \rho \sqrt{\kappa}} \int_{\Omega} |\psi(x) |^2 \, dx\;.
\end{align}
Here $K(s)$ is the function defined in \eqref{eq:Ks}.
\end{thm}

The proof of Theorem~\ref{thm:nucleation} will also be given in Section~\ref{sec:GenHC3}.

$\,$

\subsection{Discussion of critical fields}~\\
Actually, we should define more than one critical field, instead of just $H_{C_3}$.
We define an upper and a lower critical field, 
$\underline{H_{C_3}(\kappa)} \leq \overline{H_{C_3}(\kappa)}$, by
\begin{align}
\overline{H_{C_3}(\kappa)} &=  \inf\{ H>0 \;:\; \text{for all } H'>H, (0, \vec{F}) \text{ is the only minimizer of } {\mathcal E}_{\kappa,H'}\} \;,\nonumber\\
\underline{H_{C_3}(\kappa)} &= H_{C_3}(\kappa)\;.
\end{align}

The proof of Theorem~\ref{thm:GeneralHG3} gives a lower bound to $\underline{H_{C_3}(\kappa)} $ and an upper bound to $\overline{H_{C_3}(\kappa)}$, so the expansion in \eqref{eq:GenAsymp} is valid for both fields.
The physical idea of a sharp value for the external magnetic field strength at which superconductivity disappears, requires the different definitions of the critical field to coincide. Our most precise result, Theorem~\ref{thm:HPimproved}, establishes this identification under a (generically satisfied) geometric assumption on $\partial \Omega$.

Most works analyzing $H_{C_3}$ relate (more or less implicitly) these {\it global} critical fields to local ones given purely in terms of spectral data of a magnetic Schr\"{o}dinger operator, i.e. in terms of a {\it linear} problem. We will discuss the local fields more in Section~\ref{LocalFields}. Here we will give the following definition. Let, for $B \in {\mathbb R}_{+}$,  the magnetic Neumann Laplacian ${\mathcal H}(B)$ be the self-adjoint operator (with Neumann boundary conditions) associated to the  quadratic form
\begin{align}
\label{eq:Form}
W^{1,2}(\Omega) \ni u \mapsto \int_{\Omega} | (-i\nabla - B\vec{F}) u |^2\,dx\;,
\end{align}
We define $\lambda_1(B)$ as the lowest eigenvalue of ${\mathcal H}(B)$. The local fields can now be defined as follows.
\begin{align}
\label{eq:SpecLocalDef}
\overline{H_{C_3}^{\rm loc}(\kappa)} &=  \inf\{ H>0 \;:\; \text{ for all } H'>H, \lambda_1(\kappa H') \geq \kappa^2 \} \;, \nonumber \\
\underline{H_{C_3}^{\rm loc}(\kappa)} &=  \inf\{ H>0 \;:\;  \lambda_1(\kappa H) \geq \kappa^2 \}\;.
\end{align}

The difference between $\overline{H_{C_3}^{\rm loc}(\kappa)}$ and $\underline{H_{C_3}^{\rm loc}(\kappa)}$---and also between $\overline{H_{C_3}(\kappa)}$ and $\underline{H_{C_3}(\kappa)}$---can be retraced to the general non-existence of an inverse to the function $B \mapsto \lambda_1(B)$, i.e. to lack of strict monotonicity of $\lambda_1$.

\begin{remark}~\\
The detailed spectral analysis in Bauman-Phillips-Tang \cite{BPT} in the case where $\Omega$ is a disc 
does not exclude that, in this case, $\overline{H_{C_3}^{\rm loc}(\kappa)}$ and $\underline{H_{C_3}^{\rm loc}(\kappa)}$ differ even for large values of $\kappa$. They prove the estimate \cite[Theorem~7.2]{BPT},
\begin{align*}
\Big| \overline{H_{C_3}^{\rm loc}(\kappa)} - \underline{H_{C_3}^{\rm loc}(\kappa)}\Big| \leq \frac{C}{\kappa} \;,\quad\quad
\text{ in the case of the disc.}
\end{align*}
However, in Subsection~\ref{disc} we will make a precise analysis in this special case and conclude that actually (for the disc) $\overline{H_{C_3}^{\rm loc}(\kappa)} = \underline{H_{C_3}^{\rm loc}(\kappa)}$ for sufficiently large values of $\kappa$.
\end{remark}

\begin{thm}
\label{thm:LetHalvdel}~\\
Let $\Omega$ be a bounded simply-connected domain in ${\mathbb R}^2$ with smooth boundary and let $\kappa>0$, then the following general relations hold
\begin{align}
\label{first} \overline{H_{C_3}(\kappa)}&\geq \overline{H_{C_3}^{\rm loc}(\kappa)}\;, \\
\label{second} \underline{H_{C_3}(\kappa)} &\geq \underline{H_{C_3}^{\rm loc}(\kappa)}\;.
\end{align}
\end{thm}

The easy proof of 
Theorem~\ref{thm:LetHalvdel} is given in Section~\ref{LocalFields}.
For general domains we do not know that the local fields $\underline{H_{C_3}^{{\rm loc}}(\kappa)}$ and $\overline{H_{C_3}^{{\rm loc}}(\kappa)}$ coincide.

The next theorem improves Theorem \ref{thm:LetHalvdel} and is typical of type II materials.

\begin{thm}
\label{thm:Identical}~\\
Let $\Omega$ be a bounded simply-connected domain in ${\mathbb R}^2$ with smooth boundary. Then there exists a constant $\kappa_0>0$ such that, for $\kappa > \kappa_0$,  we have
\begin{align}
\overline{H_{C_3}(\kappa)} = \overline{H_{C_3}^{{\rm loc}}(\kappa)}\;.
\end{align}
\end{thm}
Theorem~\ref{thm:Identical} will be proved in Section~\ref{equal}.

$\,$

\subsection{Results for non-degenerate domains}~\\
In order to obtain more precise results, we need to impose geometric conditions on $\Omega$. We will work with two different conditions (one more strict than the other).

\begin{assumption}
\label{assump:GenNondegen}~\\
The domain $\Omega \subset {\mathbb R}^2$ is bounded and simply-connected and has smooth boundary. Furthermore, there exists a finite number of points $\{s_1, \ldots, s_N \} \in {\mathbb R}/|\partial \Omega|$ of maximal curvature, i.e. such that 
\begin{align}
&k(s_j) = \sup_{s \in {\mathbb R}/|\partial \Omega|} k(s)\;, \nonumber \\
&k(s) < k(s_j)\;, \quad \forall s \in {\mathbb R}/|\partial \Omega| \setminus \{s_1, \ldots, s_N \}\;.
\end{align}
Finally, these maxima are non-degenerate, in the sense that
$k_{2,j} := -k''(s_j) \neq 0$. We write $k_2 = \min_j k_{2,j}$.
\end{assumption}

\begin{assumption}
\label{assump:omega}~\\
The domain $\Omega \subset {\mathbb R}^2$ is bounded and simply-connected and has smooth boundary. Furthermore, there exists a unique point $s_0 \in {\mathbb R}/|\partial \Omega|$ of maximal curvature, i.e. such that $k(s_0) = \sup_{s \in {\mathbb R}/|\partial \Omega|} k(s)$, and this maximum is non-degenerate, in the sense that
$k_2 := -k''(s_0) \neq 0$.
\end{assumption}

In Fournais-Helffer \cite{FournaisHelffer2} the asymptotics of  $\lambda_1(B)$, for large $B$, was calculated (under Assumption~\ref{assump:omega}).
In Appendix~\ref{AppGenNonDeg} we prove that a similar asymptotics holds under the (less restrictive)
Assumption~\ref{assump:GenNondegen}.
We will prove in Section~\ref{diamag} that (under Assumption~\ref{assump:GenNondegen}) $\lambda_1 : [B_0, \infty) \rightarrow [\lambda_1(B_0), \infty)$ is bijective for $B_0$ sufficiently large. 
In particular, we get~:
\begin{prop}
\label{prop:InverseFunction}~\\
Suppose $\Omega$ satisfies Assumption~\ref{assump:GenNondegen}. Then there exists $\kappa_0$ such that, if $\kappa \geq \kappa_0$, then the equation for $H$:
\begin{align}
\label{eq:FormalField}
\lambda_1(\kappa H) = \kappa^2\;,
\end{align}
has a unique solution $H(\kappa)$. 
\end{prop}
In other words, Proposition~\ref{prop:InverseFunction} says that, for large $\kappa$, the upper and lower {\it local} fields, defined in \eqref{eq:SpecLocalDef}, coincide.
We define, for $\kappa \geq \kappa_0$, the local critical field $H_{C_3}^{{\rm loc}}(\kappa)$ to be the solution given by Proposition~\ref{prop:InverseFunction}, i.e.
\begin{align}
\label{eq:Hnobar}
\lambda_1(\kappa H_{C_3}^{{\rm loc}}(\kappa)) = \kappa^2\;.
\end{align}

In Subsection~\ref{CalcAsymp}, we calculate the asymptotics of $H_{C_3}^{{\rm loc}}(\kappa)$ (based on the asymptotics of $\lambda_1(B)$ from \cite{FournaisHelffer2}). The result is that the solution $H_{C_3}^{{\rm loc}}(\kappa)$ to \eqref{eq:FormalField} has the formal asymptotic expansion given by $H_{\rm formal}$, where 
\begin{align}
\label{Hformal}
H_{\rm formal} = \frac{\kappa}{\Theta_0} \Big( 1 + \frac{C_1 k_{\rm max}}{\sqrt{\Theta_0} \kappa} - C_1 \sqrt{\tfrac{3k_2}{2}} \kappa^{-\frac{3}{2}} + \kappa^{-\frac{7}{4}} \sum_{j=0}^{\infty} \eta_j \kappa^{-\frac{j}{4}}\Big)\;,
\end{align}
as $\kappa \rightarrow +\infty$.
Here the coefficients $\eta_j \in {\mathbb R}$ are computable recursively. The expression for $H_{\rm formal}$ is to be understood as an asymptotic series, no convergence being proved (or even expected).

Using Proposition~\ref{prop:InverseFunction} we can identify the lower and upper local fields and therefore find the following result.

\begin{thm}
\label{thm:HPimproved}~\\
Suppose $\Omega$ is either the disc or that it satisfies Assumption~\ref{assump:GenNondegen}. 
Then there exists $\kappa_0 >0$ such that, when $\kappa>\kappa_0$, then
\begin{align}
H_{C_3}^{{\rm loc}}(\kappa) = \underline{H_{C_3}(\kappa)} =
\overline{H_{C_3}(\kappa)}\;.
\end{align}
\end{thm}

\begin{proof}~\\
The case of the disc follows from Theorems~\ref{thm:LetHalvdel} and~\ref{thm:Identical} together with Corollary~\ref{cor:EqualDisc}. \\
For the non-degenerate case---i.e. under Assumption~\ref{assump:GenNondegen}---Theorem~\ref{thm:HPimproved} follows from combining Proposition~\ref{prop:InverseFunction} with
Theorems~\ref{thm:LetHalvdel} and~\ref{thm:Identical}.
\end{proof}

\begin{remark}~\\
Under Assumption~\ref{assump:GenNondegen}, the known asymptotics, \eqref{Hformal}, of $H_{C_3}^{{\rm loc}}(\kappa)$ can, of course, be combined with Theorem~\ref{thm:HPimproved} to find the leading order terms of the expansion of $H_{C_3}(\kappa)$ for $\kappa$ large.
\end{remark}

\section{Diamagnetism} \label{diamag}
\subsection{General domains} \label{DiaGen} ~\\
In this section we will study the behavior for large $B$ of the lowest Neumann eigenvalue $\lambda_1(B)$ of the operator ${\mathcal H}(B)$ associated to the quadratic form in \eqref{eq:Form}.

We will not generally impose the Assumption~\ref{assump:omega} in this section.
We only assume that $\Omega$ is bounded with piecewise Lipschitz boundary. Then the magnetic operator ${\mathcal H}(B)$ has compact resolvent, so the eigenvalues tend to infinity, in particular, the degeneracy of the ground state is finite.

Let $B \in {\mathbb R}$ and let $n$ be the degeneracy of $\lambda_1(B)$.
By analytic perturbation theory (see for instance \cite{Kato} or \cite[Chapter XII]{ReSi}) there exists $\epsilon >0$, $n$ analytic functions $(B-\epsilon, B+\epsilon) \ni \beta \mapsto \phi_{j}(\beta) \in H^2(\Omega)\setminus \{0\}$, 
and $n$ analytic functions $(B-\epsilon, B+\epsilon) \ni \beta \mapsto E_j(\beta) \in {\mathbb R}$,
such that 
\begin{align*}
{\mathcal H}(\beta) \phi_j(\beta) &= E_j(\beta) \phi_j(\beta)\;, &
E_j(B) &= \lambda_1(B)\;.
\end{align*}
We may choose $\epsilon$ sufficiently small in order to have the existence (but not necessarily the uniqueness) of $j_{+}, j_{-} \in \{1, \ldots, n\}$ such that
\begin{align}
&\text{For $\beta>B$: }\quad\quad\quad E_{j_+}(\beta) = \min_{j  \in \{1, \ldots, n\}} E_j(\beta)\nonumber\\
&\text{For $\beta<B$: }\quad\quad\quad E_{j_-}(\beta) = \min_{j  \in \{1, \ldots, n\}} E_j(\beta)\;.
\end{align}
Define the left and right derivatives of $\lambda_1(B)$:
\begin{align}
\lambda_{1,\pm}'(B) := \lim_{\epsilon \rightarrow 0_{\pm}} \frac{\lambda_1(B+\epsilon) - \lambda_1(B)}{\epsilon}\;. 
\end{align}
\begin{prop}
\label{prop:halfder}~\\
For all $B \in {\mathbb R}$, the one-sided derivatives $\lambda_{1,\pm}'(B)$ exist and satisfy
$$
\lambda_{1,\pm}'(B) = -2 \Re \langle \phi_{j_{\pm}} \,| \, \vec{F}\cdot (-i\nabla - B \vec{F})  \phi_{j_{\pm}}\rangle\;.
$$
\end{prop}

\begin{proof}~\\
Clearly, $\lambda_{1,\pm}'(B) = E_{j_{\pm}}'(B)$.
We will prove that
$$
E_{j_{\pm}}'(B)= -2 \Re \langle \phi_{j_{\pm}} \, | \, \vec{F}\cdot (-i\nabla - B \vec{F})  \phi_{j_{\pm}}\rangle\;.
$$
But this result is just first order perturbation theory (Feynman-Hellmann).
\end{proof}

\begin{prop}
\label{prop:limDer}~\\
Let $g$ be a function such that for all $\epsilon \in (-1,1)$ we have
\begin{align}
\label{eq:abstract}
|g(\beta+ \epsilon) - g(\beta) | \rightarrow 0\,,
\end{align}
as $\beta \rightarrow \infty$.

Suppose $\Omega$ is such that there exists $\alpha \in {\mathbb R}$ such that $\lambda_1(B) = \alpha B +g(B) + o(1)$, as $B \rightarrow + \infty$. Then the limits $\lim_{B \rightarrow \infty} \lambda_{1,+}'(B)$ and $\lim_{B \rightarrow \infty} \lambda_{1,-}'(B)$ exist and
\begin{align}
\lim_{B \rightarrow \infty} \lambda_{1,+}'(B)= 
\lim_{B \rightarrow \infty} \lambda_{1,-}'(B)
= \alpha\;.
\end{align}
\end{prop}

\begin{remark}
\label{rem:powers}~\\
Let $\gamma \in [0,1)$, then $g(\beta) = \beta^{\gamma}$ satisfies \eqref{eq:abstract}. Thus, if there exist $\gamma_1, \ldots, \gamma_m \in [0,1)$ and $\alpha,\alpha_1, \ldots, \alpha_m \in {\mathbb R}$, such that, as $B \rightarrow \infty$,
$$
\lambda_1(B) = \alpha B + \sum_{j=1}^m \alpha_j B^{\gamma_j} + o(1)\,,
$$
then Proposition~\ref{prop:limDer} implies that
\begin{align*}
\lim_{B \rightarrow \infty} \lambda_{1,\pm}'(B)
= \alpha\;.
\end{align*}
\end{remark}

\begin{proof}[Proof of Proposition~\ref{prop:limDer}]~\\
Clearly\footnote{We use here the fact that $\lambda_{j_{\pm}}$ is an analytic choice of the eigenvalues in a neighborhood of $B$.}, for all $B$, we have $\lambda_{1,+}'(B) \leq \lambda_{1,-}'(B)$. So it suffices to prove that
\begin{align}
\label{alphaunder}
&\alpha  \leq \liminf_{B \rightarrow \infty} \lambda_{1,+}'(B)\;, \\
\label{alphaover}
&\limsup_{B \rightarrow \infty} \lambda_{1,-}'(B) \leq \alpha\;.
\end{align}
We now observe that, for any $\epsilon>0$,
\begin{align*}
\lambda_{1,+}'(B) &= -2 \Re \langle \phi_{j_{+}}(B) \, | \, \vec{F}\cdot (-i\nabla - B \vec{F}) \phi_{j_{+}}(B) \rangle \\
&= \frac{1}{\epsilon} \langle \phi_{j_{+}}(B) \, | \, \big( {\mathcal H}(B+\epsilon) - {\mathcal H}(B) - \epsilon^2 \vec{F}^2\big) \phi_{j_{+}}(B) \rangle\;.
\end{align*}
Therefore, the variational principle implies
\begin{align*}
\lambda_{1,+}'(B) \geq \frac{\lambda_1(B+\epsilon) - \lambda_1(B)}{\epsilon} - \epsilon \| \vec{F} \|_{L^{\infty}(\Omega)}^2\;.
\end{align*}
By assumption there exists a function $f : {\mathbb R}_{+}  \rightarrow {\mathbb R}_{+}$, with
$\lim_{\beta \rightarrow \infty} f(\beta) = 0$, and such that
\begin{align*}
| \lambda_1(\beta) - (\alpha \beta + g(\beta)) | \leq f(\beta)\;.
\end{align*}
Thus,
\begin{align}
\label{eq:derivative}
\lambda_{1,+}'(B) \geq \alpha + \frac{g(B+\epsilon)-g(B)}{\epsilon} - \frac{f(B) + f(B+\epsilon)}{\epsilon} - \epsilon \| \vec{F} \|_{L^{\infty}(\Omega)}^2\;.
\end{align}
Therefore, (using \eqref{eq:abstract})
\begin{align*}
\liminf_{B \rightarrow \infty} \lambda_{1,+}'(B) \geq \alpha - 
\epsilon \| \vec{F} \|_{L^{\infty}(\Omega)}^2\;.
\end{align*}
Since $\epsilon > 0$ was arbitrary, this finishes the proof of \eqref{alphaunder}.

The proof of \eqref{alphaover} is similar (taking $\epsilon <0$ reverses the inequalities) and will be omitted.
\end{proof}
From Remark~\ref{rem:powers} it is clear that in order to prove monotonicity of $\lambda_1(B)$ we only need to have an asymptotic expansion of $\lambda_1(B)$ with an error term of order $o(1)$. This can be achieved in much more general situations than that given by Assumption~\ref{assump:omega}, which was the assumption used in \cite{FournaisHelffer2}. In Appendix~\ref{AppGenNonDeg} we generalize the result of \cite{FournaisHelffer2} to the situation where the curvature of the boundary has a finite number of non-degenerate maxima. In particular, we get that Proposition~\ref{prop:diamagOmega} below holds in this case.
Notice that with several maxima (and symmetry) one expects (see \cite{BoDa}) the difference between $\lambda_1(B)$ and $\lambda_2(B)$ to be exponentially small, so it may seem a bit surprising that one is able to prove Proposition~\ref{prop:diamagOmega} in this case. 

Another very interesting case where the conditions of Proposition~\ref{prop:limDer} can be verified is the case of a domain with corners. This is the subject of the work \cite{BoDa}.

\begin{prop}
\label{prop:diamagOmega}~\\
Suppose $\Omega$ satisfies
Assumption~\ref{assump:GenNondegen}.
Then 
\begin{align}
\lim_{B \rightarrow \infty} \lambda_{1,+}'(B)= 
\lim_{B \rightarrow \infty} \lambda_{1,-}'(B)
= \Theta_0\;.
\end{align}
In particular, $B \mapsto \lambda_1(B)$ is strictly increasing for large $B$.
\end{prop}

\begin{proof}~\\
This is clear using Proposition~\ref{prop:limDer}, Corollary~\ref{cor:LargeB} and Remark~\ref{rem:powers}.
\end{proof}
We finish this subsection by giving the proof of Proposition~\ref{prop:InverseFunction}.

\begin{proof}[Proof of Proposition~\ref{prop:InverseFunction}]~\\
Since, by Proposition~\ref{prop:diamagOmega}, 
$\lim_{B \rightarrow \infty} \lambda_{1,+}'(B)= 
\lim_{B \rightarrow \infty} \lambda_{1,-}'(B)= \Theta_0 >0$, there exists $B_0 >0$ such that
$B \mapsto \lambda_1(B)$ is strictly increasing from $[B_0, + \infty)$ to $[\lambda_1(B_0), + \infty)$. Furthermore, by continuity, we may choose $B_0$ sufficiently big such that
\begin{align}
\label{eq:monotoneEnd}
\lambda_1(B) < \lambda_1(B_0)\;,
\end{align}
for all $B < B_0$.

So, using \eqref{eq:monotoneEnd}, the inverse function $\lambda_1^{-1}$ is uniquely defined as a continuous function
$$
\lambda_1^{-1} : [\lambda_1(B_0), + \infty) \rightarrow [B_0, + \infty)\;.
$$
Define $\kappa_0$ by $ \kappa_0 = \sqrt{\lambda_1(B_0)}$. Then, for $\kappa> \kappa_0$, the equation
$$
\lambda_1(\kappa H) = \kappa^2\;,
$$
has the unique solution $H= \frac{\lambda_1^{-1}(\kappa^2)}{\kappa}$. 
\end{proof}

\subsection{The case of the disc}\label{disc}~\\
In the case where $\Omega=B(0,R)$ is a disc, we do not know from the best available asymptotics (\cite{BPT}) that the hypothesis \eqref{eq:abstract} is satisfied.
In this subsection we will make a more precise asymptotic estimate in order to settle the question of diamagnetism for the disc. 
The definition of the spectral parameters, $C_1, \Theta_0, \xi_0$ is recalled in Appendix~\ref{AppA}.

\begin{thm}[Eigenvalue asymptotics for the disc]
\label{thm:disc}~\\
Suppose that $\Omega$ is the unit disc. 
Define $\delta(m,B)$, for $m\in {\mathbb Z}$, $B>0$, by
\begin{align}
\label{eq:delta}
\delta(m,B) = m - \tfrac{B}{2} - \xi_0 \sqrt{B}.
\end{align}
Then there exist (computable) constants $C_0, \delta_0 \in {\mathbb R}$ such that if
\begin{align}
\Delta_B = \inf_{m \in {\mathbb Z}} | \delta(m,B) - \delta_0|\;,
\end{align}
then
\begin{align}
\lambda_1(B) = \Theta_0 B - C_1 \sqrt{B} +
3C_1 \sqrt{\Theta_0} \big( \Delta_B^2 + C_0\big) + {\mathcal O}(B^{-\frac{1}{2}})\;.
\end{align}
\end{thm}

\begin{remark}~\\
As the proof will show, the constants $C_0, \delta_0$ can be expressed in terms of spectral data for the basic operator ${\mathfrak h}_0$ (see \eqref{eq:hs}) discussed in Appendix~\ref{AppA}.
\end{remark}

Before we give the proof of Theorem~\ref{thm:disc}, we collect the following important consequence.

\begin{prop}
\label{prop:DiamagDisc}~\\
Let $\Omega$ be the disc. Then the left- and right-hand derivatives $\lambda_{1,\pm}'(B)$ exist and satisfy
\begin{align}
&\lambda_{1,+}'(B) \leq  \lambda_{1,-}'(B)\;, \nonumber\\
&\liminf_{B \rightarrow + \infty} \lambda_{1,+}'(B) \geq
\Theta_0 - \tfrac{3}{2}C_1 |\xi_0| >0\;.
\end{align}
In particular, $B \mapsto \lambda_1(B)$ is strictly increasing for large $B$.
\end{prop}

The monotonicity of $B \mapsto \lambda_1(B)$ implies that the local fields are equal.
\begin{cor}
\label{cor:EqualDisc}~\\
Let $\Omega$ be the disc. Then there exists a constant $\kappa_0>0$ such that if
$\kappa>\kappa_0$ then
\begin{align}
\underline{H_{C_3}^{{\rm loc}}(\kappa)}
=\overline{H_{C_3}^{{\rm loc}}(\kappa)}\;.
\end{align}
\end{cor}

\begin{proof}[Proof of Proposition~\ref{prop:DiamagDisc}]~\\
Let $g(B) = - C_1 \sqrt{B} +
3C_1 \sqrt{\Theta_0} \big( \Delta_B^2 + C_0\big)$, $\alpha = \Theta_0$.
We calculate as in the proof of Proposition~\ref{prop:limDer} until we reach \eqref{eq:derivative}.
Notice that $0 \leq \Delta_B \leq \frac{1}{2}$, for all $B>0$.
Furthermore, consider $B>1$, $\epsilon>0$. Let $m_0\in {\mathbb Z}$ be such that
$\Delta_{B+\epsilon} = |m_0 - \tfrac{B+\epsilon}{2} - \xi_0 \sqrt{B+\epsilon}-\delta_0|$. Then, since $-1<\xi_0 <0$, 
\begin{align}
\Delta_{B+\epsilon} - \Delta_B 
&\geq 
\big|m_0 - \frac{B+\epsilon}{2} - \xi_0 \sqrt{B+\epsilon}-\delta_0\big|
-\big|m_0 - \frac{B}{2} - \xi_0 \sqrt{B}-\delta_0\big| \nonumber\\
&\geq
-\big|- \frac{\epsilon}{2} - \xi_0 \frac{\epsilon}{\sqrt{B+\epsilon}+\sqrt{B}}\big|
\geq -\frac{\epsilon}{2}.
\end{align}
Therefore,
$$
\Delta_{B+\epsilon}^2 - \Delta_B^2 = (\Delta_{B+\epsilon} + \Delta_B)(\Delta_{B+\epsilon} - \Delta_B)
\geq -\frac{\epsilon}{2}\;,
$$
and we get 
$$
\liminf_{B \rightarrow +\infty} \frac{g(B+\epsilon) - g(B)}{\epsilon} \geq - \tfrac{3}{2} C_1 \sqrt{\Theta_0}\;.
$$
The rest of the proof follows the one of Proposition~\ref{prop:limDer} by taking $\epsilon$ to zero (and using that $|\xi_0| = \sqrt{\Theta_0}$).

That $\Theta_0 > \frac{3}{2}C_1 |\xi_0|$ can be seen from the following argument.
From \cite[Proposition~A.3]{FournaisHelffer2}, we get that $3C_1 |\xi_0| = 1-4I_2$, where the integral $I_2$ (given in \eqref{eq:I2} below) satisfies $I_2>0$. In particular, $3C_1 |\xi_0|<1$. 
Since it is known that $\Theta_0> \frac{1}{2}$, this proves the desired statement.

We also state the following numerical values from \cite{Bo1},
\begin{align}
C_1 &= 0.254\;, & |\xi_0| &=0.768\;.
\end{align}
\end{proof}

\begin{proof}[Proof of Theorem~\ref{thm:disc}]~\\
We will use standard results on the spectral theory of the Neumann operator
 $$h(\zeta) = -\frac{d^2}{d\tau^2} + (\tau + \zeta)^2$$ on $L^2({\mathbb R}_{+},d\tau)\;.$ Some of these results are recalled in Appendix~\ref{AppA}.

Let $D(t) = \{ x \in {\mathbb R}^2 \,| \, |x| \leq t\}$ be the disc with radius $t$.
Let $\widetilde{Q}_B$ be the quadratic form
$$
\widetilde{Q}_B[u] = \int_{D(1) \setminus D(\frac{1}{2})} \big|(-i\nabla -B\vec{F})u\big|^2\,dx\;,
$$
with domain $\{u \in H^1(D(1) \setminus D(\frac{1}{2})) \,| \, u(x) = 0 \text{ on } |x|=\frac{1}{2} \}$.
Let $\tilde{\lambda}(B)$ be the lowest eigenvalue of the corresponding self-adjoint operator (Friedrichs extension).
Using the Agmon estimates in the normal direction (see Theorem~\ref{thm:agmon}), we see that
\begin{align}
\label{eq:trekant}
\lambda_1(B) = \tilde{\lambda}(B) + {\mathcal O}(B^{-\infty})\;.
\end{align}

By changing to boundary coordinates (if $(r,\theta)$ are usual polar coordinates, then $t=1-r$, $s=\theta$), the quadratic form $\widetilde{Q}_B[u]$ becomes,
\begin{align}
\widetilde{Q}_B[u] &= \int_0^{2\pi} ds \int_0^{1/2} dt \,(1-t)^{-1} | (D_s - B \tilde{A}_1)u|^2+
(1-t) |D_t u|^2\;, \\
\| u \|_{L^2}^2 &=  \int_0^{2\pi} ds \int_0^{1/2} dt \,(1-t) |u|^2\;,
\quad \tilde{A}_1 = \tfrac{1}{2} - t + \tfrac{t^2}{2}\;.\nonumber
\end{align}
Here we used Lemma~\ref{lem:GoodGaugeImp}, and that 
$\gamma_0 = \frac{\int_{\Omega}\curl \vec{F}\,dx}{|\partial \Omega|} = \frac{1}{2}$ for the disc.

Performing the scaling $\tau = \sqrt{B} t$ and decomposing in Fourier modes, 
$$
u = \sum_m e^{im s} \phi_m(t)\;,
$$ 
we find
\begin{align}
\label{eq:stjerne}
\tilde{\lambda}(B) = B \inf_{m \in {\mathbb Z}} e_{\delta(m,B), B}\;.
\end{align}
Here the function $\delta(m, B)$ was defined in \eqref{eq:delta} 
and $e_{\delta,B}$ is the lowest eigenvalue of the quadratic form $q_{\delta,B}$ on 
$L^2((0, \sqrt{B}/2);(1-\sqrt{B}\tau)d\tau)$ (with Neumann boundary condition at $0$ and Dirichlet at $\sqrt{B}/2$):
\begin{align}
q_{\delta,B}[\phi] = \int_0^{\sqrt{B}/2} &(1-\tfrac{\tau}{\sqrt{B}})^{-1} \big( (\tau+\xi_0) +B^{-\frac{1}{2}}(\delta -\tfrac{\tau^2}{2}) \big)^2 \nonumber\\
&+ (1-\tfrac{\tau}{\sqrt{B}}) |\phi'(\tau)|^2\,d\tau \;.
\end{align}
We will only consider $\delta$ varying in a fixed bounded set. This is justified since it follows from \cite[Lemma 5.4]{FournaisHelffer2} that for all $C>0$ there exists $D>0$ such that if $|\delta| > D$ and $B>D$, then 
$$
e_{\delta,B} \geq \Theta_0 - C_1 B^{-\frac{1}{2}} + C\;.
$$

Furthermore, for $\delta$ varying in a fixed bounded set, we know (from the analysis of ${\mathfrak h}_0$, some of which is recalled in Appendix~\ref{AppA}) that there exists a $d>0$ such that if $B>d^{-1}$, then
the spectrum of $q_{\delta,B}$ contained in $(-\infty, \Theta_0+d)$ consists of exactly one simple eigenvalue.

The self-adjoint Neumann operator ${\mathfrak h}(\delta,B)$ associated to
$q_{\delta,B}$ (on the space $L^2((0, \sqrt{B}/2);(1-\sqrt{B}\tau)d\tau)$) is
\begin{align}
{\mathfrak h}(\delta,B)&=
-(1-\tfrac{\tau}{\sqrt{B}})^{-1} \frac{d}{d\tau} (1-\tfrac{\tau}{\sqrt{B}}) \frac{d}{d\tau}\nonumber\\
&\quad+(1-\tfrac{\tau}{\sqrt{B}})^{-2} \big( (\tau+\xi_0) +B^{-\frac{1}{2}}(\delta -\tfrac{\tau^2}{2}) \big)^2\;.
\end{align}
We will write down an explicit test function for ${\mathfrak h}(\delta,B)$ in \eqref{eq:trial} below, giving $e_{\delta,B}$ up to an error of order ${\mathcal O}(B^{-\frac{3}{2}})$ (locally uniformly in $\delta$).

We can formally develop ${\mathfrak h}(\delta,B)$ as
$$
{\mathfrak h}(\delta,B) = {\mathfrak h}_0 + B^{-\frac{1}{2}} {\mathfrak h}_1 + B^{-1} {\mathfrak h}_2 + {\mathcal O}(B^{-\frac{3}{2}})\;.
$$
with
\begin{align}
\label{eq:hs}
{\mathfrak h}_0 &= -\frac{d^2}{d\tau^2} + (\tau+\xi_0)^2\;, \nonumber\\
{\mathfrak h}_1 &=\frac{d}{d\tau} + 2 (\tau+\xi_0) (\delta -\tfrac{\tau^2}{2}) + 2\tau (\tau+\xi_0)\;,\nonumber\\
{\mathfrak h}_2 &= \tau\frac{d}{d\tau} +  (\delta -\tfrac{\tau^2}{2})^2 + 4 \tau (\tau+\xi_0) (\delta -\tfrac{\tau^2}{2})
+ 3 \tau^2 (\tau+\xi_0)^2\;.
\end{align}
Let $u_0$ be the known ground state eigenfunction of ${\mathfrak h}_0$ with eigenvalue $\Theta_0$. Here ${\mathfrak h}_0$ is considered as a selfadjoint operator on $L^2({\mathbb R}_{+}; d\tau)$ with Neumann boundary condition at $0$.
Let $R_0$ be the regularized resolvent, which is defined by
$$
R_0 \phi = \begin{cases}({\mathfrak h}_0 -\Theta_0)^{-1} \phi\;, & \phi \perp u_0\;, \\ \quad 0\;, &
\phi \parallel u_0\,.
\end{cases}
$$
Here `perpendicular' is measured with respect to the usual inner product (no perturbation of the measure) in $L^2({\mathbb R}_{+}; d\tau)$.

Let $\lambda_1$ and $\lambda_2$ be given by
\begin{align}
\lambda_1 &:= \langle u_0 \,|\,{\mathfrak h}_1 u_0 \rangle\;, &
\lambda_2 &:= \lambda_{2,1} + \lambda_{2,2}\;,\nonumber\\
\lambda_{2,1}&:= \langle u_0 \,|\,{\mathfrak h}_2 u_0 \rangle \;,&
\lambda_{2,2}&:= \langle u_0 \,|\,({\mathfrak h}_1 - \lambda_1) u_1 \rangle \;,
\end{align}
Here the inner products are the usual inner products in $L^2({\mathbb R}_{+}; d\tau)$.
The functions $u_1, u_2$ are given as
\begin{align}
u_1 &= - R_0 ({\mathfrak h}_1 - \lambda_1) u_0\;,
&
u_2 &= - R_0 \big\{ ({\mathfrak h}_1 - \lambda_1) u_1 + ({\mathfrak h}_2 - \lambda_2)u_0 \big\}\;.
\end{align}
Notice that (see \cite[Lemma~A.5]{FournaisHelffer2}) $u_0 \in {\mathcal S}(\overline{{\mathbb R}_{+}})$ and that $R_0$ maps ${\mathcal S}(\overline{{\mathbb R}_{+}})$ (continuously) into itself.
Therefore, $u_0, u_1, u_2$ (and their derivatives) are rapidly decreasing functions on ${\mathbb R}_{+}$.

Let $\chi \in C_0^{\infty}({\mathbb R})$ be a usual cut-off function, such that
\begin{align}
\chi(t) &= 1 \quad \text{ for } |t|\leq \tfrac{1}{8} \;, &
\supp \chi &\subset [-\tfrac{1}{4}, \tfrac{1}{4}]\;,
\end{align}
and let $\chi_B(\tau) = \chi(\tau B^{-\frac{1}{4}})$\,.

Our trial state is defined by
\begin{align}
\label{eq:trial}
\psi := \chi_B \big\{ u_0 + B^{-\frac{1}{2}} u_1 + B^{-1} u_2 \big\}\;.
\end{align}
A calculation (using in particular the exponential decay of the involved functions) gives that
\begin{align}
&\big\| \big\{{\mathfrak h}(\delta,B) - \big(\Theta_0 + \lambda_1 B^{-\frac{1}{2}} + \lambda_2 B^{-1}\big)\big\}
\psi \big\|_{L^2([0, \sqrt{B}/2];(1-\sqrt{B}\tau)d\tau)} = {\mathcal O}(B^{-\frac{3}{2}})\;,\nonumber\\
&\| \psi \|_{L^2([0, \sqrt{B}/2];(1-\sqrt{B}\tau)d\tau)} = 1 +{\mathcal O}(B^{-\frac{1}{2}})\;,
\end{align}
where the constant in ${\mathcal O}$ is uniform for $\delta$ in bounded sets.

Therefore, we have proved that (uniformly for $\delta$ varying in bounded sets)
\begin{align}
e_{\delta, B} = \Theta_0 + \lambda_1 B^{-\frac{1}{2}} + \lambda_2 B^{-1} +{\mathcal O}(B^{-\frac{3}{2}})\;.
\end{align}
It remains to calculate $\lambda_1, \lambda_2$ and, in particular, deduce their dependence on $\delta$.

A standard calculation (which can for instance be found in \cite[Section 2]{FournaisHelffer2}) gives that
\begin{align}
\lambda_1=-C_1\;.
\end{align}
In particular, $\lambda_1$ is independent of $\delta$ (this is because $\int_0^{\infty} (\tau + \xi_0) u_0^2 \,d\tau = 0$).

It is much harder to calculate $\lambda_2$ explicitly. However, notice (by writing 
$
\lambda_{2,2} = - \langle u_0 \, | \, ({\mathfrak h}_1 - \lambda_1) R_0 ({\mathfrak h}_1 - \lambda_1) u_0 \rangle
$) that $\lambda_2(\delta)$ is a quadratic polynomial as a function of $\delta$.
We find the coefficient to $\delta^2$ as
$ 1 - 4 I_2$, with 
\begin{align}
\label{eq:I2}
I_2 := \langle u_0 \, | \, (\tau + \xi_0) R_0 (\tau + \xi_0) u_0 \rangle\;.
\end{align}
But this integral was calculated in \cite[Proposition~A.3]{FournaisHelffer2} (we have kept the notation $I_2$ from that paper), and it was found that $1 - 4 I_2 = 3 C_1 \sqrt{\Theta_0}$.
Notice that $3 C_1 \sqrt{\Theta_0} >0$.
Since $\lambda_2$ is quadratic in $\delta$, there exist in principle computable constants $\delta_0, C_0 \in {\mathbb R}$ such that
$$
\lambda_2 = 3 C_1 \sqrt{\Theta_0}\big( (\delta - \delta_0)^2 + C_0\big)\;.
$$
Remembering \eqref{eq:trekant} and \eqref{eq:stjerne} this finishes the proof of Theorem~\ref{thm:disc}.
\end{proof}

\section{Local critical fields}
\label{LocalFields}
\subsection{General analysis}~\\
In addition to the (global) critical fields $\overline{H_{C_3}(\kappa)}$ and $\underline{H_{C_3}(\kappa)}$, one can also define local fields. 

These local fields are determined by the values where the normal solution\footnote{
Notice that $(0,\vec{F})$ is a solution to the GL-equations \eqref{eq:GL}, for all values of $\kappa,H$. Thus, $(0,\vec{F})$ is always a stationary point of the Ginzburg-Landau functional ${\mathcal E}_{\kappa, H}$\,.
} 
$(0, \vec{F})$ is a {\it not unstable} local minimum of ${\mathcal E}_{\kappa, H}$, i.e.
\begin{align}
\overline{H_{C_3}^{\rm loc}(\kappa)} &=  \inf\{ H>0 \;:\; \text{ for all } H'>H, \Hess {\mathcal E}_{\kappa, H'} \big |_{(0, \vec{F})} \geq 0 \} \;,\nonumber\\
\underline{H_{C_3}^{\rm loc}(\kappa)} &=  \inf\{ H>0 \;:\;  \Hess {\mathcal E}_{\kappa, H} \big |_{(0, \vec{F})} \geq 0 \}\;.
\end{align}
Since the Hessian, $\Hess {\mathcal E}_{\kappa, H}$, at the normal solution is associated with the sesqui\-linear form
\begin{align}
(\phi, \vec{a}) \mapsto 
\int_{\Omega} | (-i\nabla - \kappa H \vec{F}) \phi |^2 - \kappa^2 |\phi |^2
+ (\kappa H)^2 |\curl \vec{a} |^2 \,dx\;, 
\end{align}
we get the equivalent definitions given in \eqref{eq:SpecLocalDef} in the introduction. Furthermore, we get the general comparison between the local and global fields given in Theorem~\ref{thm:LetHalvdel}:

\begin{proof}[Proof of Theorem~\ref{thm:LetHalvdel}]~\\
We first prove (\ref{first}). 
Suppose $H>\overline{H_{C_3}(\kappa)}$. Then $(0,\vec{F})$ is the only minimizer of ${\mathcal E}_{\kappa, H}$. In particular, for all $\phi, \vec{A}$,
$$
{\mathcal E}_{\kappa, H}[\phi,\vec{F} + \vec{A}]
\geq {\mathcal E}_{\kappa, H}[0,\vec{F}] = 0\;.
$$
This implies that $\Hess {\mathcal E}_{\kappa, H}\big |_{(0, \vec{F})} \geq 0$\,.
Since $H>\overline{H_{C_3}(\kappa)}$ was arbitrary, we get (\ref{first}).

Next we prove (\ref{second}). 
Suppose $H < \underline{H_{C_3}^{\rm loc}(\kappa)}$. Then $\lambda_1(\kappa H) < \kappa^2$. Let $\psi$ be a ground state for ${\mathcal H}(\kappa H)$. We use, for $\eta>0$, the pair $(\eta \psi, \vec{F})$ as a trial state in ${\mathcal E}_{\kappa, H}$,
\begin{align*}
{\mathcal E}_{\kappa, H}[\eta \psi, \vec{F}] = (\lambda_1(\kappa H) - \kappa^2) \eta^2 \| \psi \|_{L^2(\Omega)}^2 + \frac{\kappa^2}{2} \eta^4 \| \psi \|_{L^4(\Omega)}^4\;.
\end{align*}
Since $\lambda_1(\kappa H) - \kappa^2<0$, we get
${\mathcal E}_{\kappa, H}[\eta \psi, \vec{F}] <0$ for $\eta$ sufficiently small (using that $W^{1,2}(\Omega) \subset L^4(\Omega)$).
Thus $(0, \vec{F})$ is not a minimizer for ${\mathcal E}_{\kappa, H}$. Since $H < \underline{H_{C_3}^{\rm loc}(\kappa)}$ was arbitrary, this proves (\ref{second}) and therefore finishes the proof of the theorem.
\end{proof}

As a corollary of Proposition~\ref{prop:InverseFunction} we get the following result.

\begin{thm}
\label{thm:EqualLocalFields}~\\
Suppose that $\Omega$ satisfies Assumption~\ref{assump:GenNondegen}.
Then there exists $\kappa_0 >0$, such that, for all $\kappa > \kappa_0$,
$\underline{H_{C_3}^{\rm loc}(\kappa)}  = \overline{H_{C_3}^{\rm loc}(\kappa)}=H_{C_3}^{\rm loc}(\kappa)$\,.
\end{thm}
\begin{proof}~\\
Proposition~\ref{prop:InverseFunction} combined with the characterization \eqref{eq:SpecLocalDef} gives the proof.
(Recall that $H_{C_3}^{\rm loc}(\kappa)$ was defined in \eqref{eq:Hnobar}).
\end{proof}

$\,$

\subsection{Calculating asymptotics}~\\
\label{CalcAsymp}
In this section we will describe the structure of the asymptotics of the solution $H(\kappa)$ to
$$
\lambda_1(\kappa H) = \kappa^2\;.
$$
The calculation is based on the asymptotic expansion of $\lambda_1(B)$ proved in
the work \cite{FournaisHelffer2} and its extension in Appendix~\ref{AppGenNonDeg}.
Therefore we need to impose Assumption~\ref{assump:GenNondegen}.

\begin{lemma}~\\
Suppose $\Omega$ satisfies Assumption~\ref{assump:GenNondegen}. Let $H=H(\kappa)$ be the solution to the equation
\begin{align}
\lambda_1(\kappa H) = \kappa^2\;,
\end{align}
given by Proposition~\ref{prop:InverseFunction}. Then there exists a sequence $\{\eta_j \}_{j=0}^{\infty} \subset {\mathbb R}$ such that
\begin{align}
\label{eq:FormalSol2}
H = \frac{\kappa}{\Theta_0} \Big( 1 + \frac{C_1 k_{\rm max}}{\sqrt{\Theta_0} \kappa} - C_1 \sqrt{\tfrac{3k_2}{2}} \kappa^{-3/2} + \kappa^{-7/4} \sum_{j=0}^{\infty} \eta_j \kappa^{-j/4}\Big)\;,
\end{align}
in the sense of an asymptotic series as $\kappa \rightarrow +\infty$\,.
\end{lemma}

\begin{proof}~\\
Let $\mu^{(1)}(h)$ be the lowest eigenvalue of the magnetic Neumann Laplacian associated with the following quadratic form on $W^{1,2}(\Omega)$
$$
\int_{\Omega} | (-ih\nabla - \vec{F}) u |^2\,dx.
$$
From \cite{FournaisHelffer2} and/or Appendix~\ref{AppGenNonDeg} we know that there exists a sequence $\{\zeta_j \}_{j=0}^{\infty} \subset {\mathbb R}$ such that for all $M \in {\mathbb N}$,
\begin{align}
\mu^{(1)}(h) = \Theta_0 h - C_1 k_{\rm max} h^{\frac{3}{2}} + C_1 \sqrt[4]{\Theta_0}\sqrt{\tfrac{3k_2}{2}}h^{\frac{7}{4}} +
h^{\frac{15}{8}} \sum_{j=0}^M \zeta_j h^{\frac{j}{8}} + {\mathcal O}(h^{2+\frac{M}{8}})\;,
\end{align}
 as $h \rightarrow 0_{+}$.
By a simple scaling we see that if $\lambda_1(B)$ is the lowest eigenvalue of the magnetic Neumann Laplacian associated to the form defined in \eqref{eq:Form},
then
$\lambda_1(B) = B^2 \mu^{(1)}(B^{-1})$, and therefore, as $B\rightarrow \infty$, we get for any $M \in {\mathbb N}$,
\begin{align}
\label{eq:Blarge}
\lambda_1(B) &= \Theta_0 B - C_1 k_{\rm max} B^{\frac{1}{2}} + C_1 \sqrt[4]{\Theta_0}\sqrt{\tfrac{3k_2}{2}}B^{\frac{1}{4}} \nonumber \\
&\quad+
B^{\frac{1}{8}} \sum_{j=0}^M \zeta_j B^{-\frac{j}{8}} + {\mathcal O}(B^{-\frac{M}{8}})\;.
\end{align}
We calculate with the {\it Ansatz} for $H(\kappa)$ given by \eqref{eq:FormalSol2}.
\begin{align}
\lambda_1&(\kappa H) \sim \kappa^2\Big( 1 + \frac{C_1 k_{\rm max}}{\sqrt{\Theta_0} \kappa} - C_1 \sqrt{\tfrac{3k_2}{2}} \kappa^{-\frac{3}{2}} + \kappa^{-\frac{7}{4}} \sum_{j=0}^{\infty} \eta_j \kappa^{-\frac{j}{4}}\Big) \\
&\quad- C_1 k_{\rm max} \frac{\kappa}{\sqrt{\Theta_0} }\Big( 1 + \frac{C_1 k_{\rm max}}{\sqrt{\Theta_0} \kappa} - C_1 \sqrt{\tfrac{3k_2}{2}} \kappa^{-\frac{3}{2}} + \kappa^{-\frac{7}{4}} \sum_{j=0}^{\infty} \eta_j \kappa^{-\frac{j}{4}}\Big)^{\frac{1}{2}} \nonumber\\
&\quad
+ C_1 \sqrt[4]{\Theta_0}\sqrt{\tfrac{3k_2}{2}}
\frac{\sqrt{\kappa}}{\sqrt[4]{\Theta_0}} \Big( 1 + \frac{C_1 k_{\rm max}}{\sqrt{\Theta_0} \kappa} - C_1 \sqrt{\tfrac{3k_2}{2}} \kappa^{-\frac{3}{2}} + \kappa^{-\frac{7}{4}} \sum_{j=0}^{\infty} \eta_j \kappa^{-\frac{j}{4}}\Big)^{\frac{1}{4}} \nonumber\\
&\quad +
\sum_{j=0}^{\infty} \zeta_j \frac{\kappa^{\frac{1-j}{4}}}{\Theta_0^{\frac{1-j}{8}}}
\Big( 1 + \frac{C_1 k_{\rm max}}{\sqrt{\Theta_0} \kappa} - C_1 \sqrt{\tfrac{3k_2}{2}} \kappa^{-\frac{3}{2}} + \kappa^{-\frac{7}{4}} \sum_{k=0}^{\infty} \eta_k \kappa^{-\frac{k}{4}}\Big)^{\frac{1-j}{8}} 
\displaybreak[0]  \nonumber\\
&=
\kappa^2 + \frac{C_1 k_{\rm max}}{\sqrt{\Theta_0}}\kappa - C_1 \sqrt{\tfrac{3k_2}{2}} \kappa^{\frac{1}{2}} + \kappa^{\frac{1}{4}} \sum_{j=0}^{\infty} \eta_j \kappa^{-\frac{j}{4}} \nonumber\\
&\quad- C_1 k_{\rm max} \frac{\kappa}{\sqrt{\Theta_0} }
\Big( 1 + \frac{1}{2}\frac{C_1 k_{\rm max}}{\sqrt{\Theta_0} \kappa} -  \frac{1}{2}C_1 \sqrt{\tfrac{3k_2}{2}} \kappa^{-\frac{3}{2}} + \kappa^{-\frac{7}{4}} \sum_{j=0}^{\infty} f_j^{(1)} \kappa^{-\frac{j}{4}}\Big)\nonumber\\
&\quad
+ C_1 \sqrt{\tfrac{3k_2}{2}} \sqrt{\kappa}
\Big( 1 + \frac{1}{4}\frac{C_1 k_{\rm max}}{\sqrt{\Theta_0} \kappa} -  \frac{1}{4}C_1 \sqrt{\tfrac{3k_2}{2}} \kappa^{-\frac{3}{2}} + \kappa^{-\frac{7}{4}} \sum_{j=0}^{\infty} f_j^{(2)} \kappa^{-\frac{j}{4}}\Big)\nonumber\\
&\quad +
\sum_{j=0}^{\infty} \zeta_j \frac{\kappa^{\frac{1-j}{4}}}{\Theta_0^{\frac{1-j}{8}}}
\Big( 1 + \frac{1-j}{8}\frac{C_1 k_{\rm max}}{\sqrt{\Theta_0} \kappa} - \frac{1-j}{8}C_1 \sqrt{\tfrac{3k_2}{2}} \kappa^{-\frac{3}{2}} + \kappa^{-\frac{7}{4}} \sum_{k=0}^{\infty} f_{j,k} \kappa^{-\frac{k}{4}}\Big) \;.\nonumber
\end{align}
Here the coefficients $f_j^{(1)}$, $f_j^{(2)}$, $f_{j,k}$ only depend on $(\eta_0, \ldots, \eta_j)$\,.
Therefore, we find the following structure
\begin{align*}
\lambda_1(\kappa H) 
\sim \kappa^2 + \kappa^{\frac{1}{4}} \sum_{j=0}^{\infty} \kappa^{-\frac{j}{4}} ( \eta_j + g_j )\;,
\end{align*}
where the $g_j$ only depend on the $\eta_s$ with $s<j$\,.
This implies that there exists a solution of the form \eqref{eq:FormalSol2} to the identity
$\lambda_1(\kappa H) = \kappa^2$
in the sense of asymptotic series.

It is elementary to prove by induction that the solution $H(\kappa)$ given by Proposition~\ref{prop:InverseFunction} must have the asymptotic expansion given by the formal solution  \eqref{eq:FormalSol2}.
\end{proof}

$\,$

\begin{remark}[Comparison with Bernoff-Sternberg~\cite{BeSt} results]~\\
In  \cite[Formula (3.1)]{BeSt} the asymptotics
$$
H^{\rm BS}_{C_3} = \overline{h} k + \frac{\overline{h}}{3J_0} \overline{\kappa} - \frac{1}{2J_0} \sqrt{\frac{-2 \overline{\kappa_{ss}} \overline{h}}{3}} \frac{1}{\sqrt{k}} + {\mathcal O}(\frac{1}{k})\,,
$$
is given. To translate to our notations we use the table
\begin{align}
k^{\rm BS}& = \kappa\,, &
\overline{h}^{\rm BS} & = \frac{1}{\Theta_0}\,, &
\overline{\kappa}^{\rm BS} &= k_{\rm max}\,, &
\overline{\kappa_{ss}}^{\rm BS}& =- k_2\,, &
J_0 &= \frac{\sqrt{\Theta_0}}{3C_1}\,.
\end{align}
So in our notation, their result is
$$
H^{\rm BS}_{C_3} = \frac{\kappa}{\Theta_0} +\frac{C_1 k_{\rm max}}{\Theta_0^{3/2}}
- \sqrt{\tfrac{3}{2}} \frac{\sqrt{k_2}}{\Theta_0} C_1 \kappa^{-1/2} + {\mathcal O}(\frac{1}{\kappa})\;.
$$
This is in almost complete agreement with our result \eqref{eq:FormalSol2}, except for the fact that we do not exclude the existence of a term of order
${\mathcal O}(\kappa^{-3/4})$\,.
\end{remark}

\section{Localization}

In this section we will give Agmon estimates for the linear problem---both tangential and normal to the boundary. We also recall how these estimates carry over to 
the non-linear equations \eqref{eq:GL}.

\subsection{Estimates in the normal direction}~\\
First we recall the estimate in the normal direction from in \cite{He-Pan} (see also the work on the linear problem \cite{He-Mo}).
Define the magnetic quadratic form by
\begin{align}
\label{eq:MagQuad}
 W^{1,2}(\Omega) \ni u \mapsto q_{B\vec{A}} = \int_{\Omega} |p_{B\vec{A}} u|^2\,dx\;,
\end{align}
where $\vec{A}$ is any (possibly $B$-dependent) vector field satisfying, for some $C_0>0$, the following estimates:
\begin{align}
\label{eq:UnifEst}
\| \vec{A} \|_{C^2(\overline{\Omega})} &\leq C_0\;, &
\| \curl \vec{A} - 1 \|_{C^1(\overline{\Omega})} &\leq C_0B^{-1/4} \;,&
\curl \vec{A} &= 1 \quad \text{on }\partial\Omega\;.
\end{align}
Notice that, by \cite[Proposition~4.2]{He-Pan} (recalled as Theorem~\ref{thm:elliptic} below), \eqref{eq:UnifEst} is verified uniformly for all the minimizers of the GL-functional.

We will denote the unique (Neumann) self-adjoint operator associated with the quadratic form $q_{B\vec{A}}$ by ${\mathcal H}_{B\vec{A}}$.

\begin{thm}[Uniform normal Agmon estimates (linear case)]
\label{thm:agmon}~\\
Let $C_0$ be given and let $\Omega$ be a bounded simply-connected domain with smooth boundary. Then there exist $C, \alpha$ such that if $\vec{A} = \vec{A}_{B}$ is a vector field satisfying \eqref{eq:UnifEst} (with the given $C_0$) and if $\phi_{B\vec{A}}$ is an eigenfunction of ${\mathcal H}(B \vec{A})$, 
with eigenvalue $\lambda = \lambda(B) \leq (1 - C_0^{-1})B$,
then,
\begin{align}
\int_{\Omega} e^{\alpha \sqrt{B} t(x)} \big( |\phi_{B\vec{A}}|^2 + B^{-1} | p_{B \vec{A}} \phi_{B\vec{A}}|^2\big)\,dx 
\leq
C \int_{\Omega}  |\phi_{B\vec{A}}|^2\,dx \;,
\end{align}
for all $B > C$.
\end{thm}
The above result can also be applied to obtain similar localization estimates for the (non-linear) Ginzburg-Landau problem. This was carried out in \cite[Proposition~4.2]{He-Pan} and~\cite[Lemma~7.2]{Pan}.
\begin{thm}[Uniform normal Agmon estimates (non-linear case)]
\label{thm:agmonNL}~\\
Let $\delta>0$.
Then there exists $\alpha, C, \kappa_0>0$ such that if 
$(\psi,\vec{A})_{\kappa, H}$ are minimizers of the Ginzburg-Landau functional, with
$(\kappa, H)$ satisfying that
\begin{align}
\kappa/H &< 1 - \delta\;,&
\kappa &> \kappa_0\;,
\end{align}
then
\begin{align}
\int_{\Omega} e^{\alpha \sqrt{\kappa H} t(x)} \big( |\psi|^2 + \tfrac{1}{\kappa H} | p_{\kappa H \vec{A}} \psi|^2\big)\,dx 
\leq
C \int_{\Omega}  |\psi|^2\,dx \;.
\end{align}
\end{thm}
This theorem admits the following but basic corollary~:
\begin{corollary}\label{CorNL}~\\
With the assumptions of the theorem, for any $p \geq 2$, there exists a constant $C_p$ such that
\begin{equation}\label{2p}
 ||\psi||_{L^2(\Omega)}
\leq C_p \kappa^{ - \frac{p-2}{2p}} ||\psi||_{L^p(\Omega)} 
\;.
\end{equation}
\end{corollary}
\begin{proof}~\\
We first observe that the normal Agmon estimate gives the existence of $C$
 such that~:
$$
 ||\psi||_{L^2(\Omega)}^2 \leq C \int_{d(x,\partial \Omega)\leq C/\kappa}
 |\psi(x)|^2 \, dx\;.
$$
We can then use H\"older to get for any $q\geq 1$
$$
 ||\psi||_{L^2(\Omega)}^2 \leq \tilde C \kappa^{- (1-\frac 1q)}
 (\int 
 |\psi(x)|^{2q} \, dx)^\frac 1q \;.
$$
Taking $q= \frac p2$, we get \eqref{2p}.
\end{proof}

\subsection{Energy estimates}\label{sec:UnifEnergy}~\\
In this subsection we will give uniform lower bounds on the ground state energies of the magnetic quadratic form $q_{B\vec{A}}$.

\begin{thm}
\label{thm:UnifLowerBounds}~\\
Let $C_0>0$ be given and let $\Omega$ be a bounded, connected domain in ${\mathbb R}^2$ with smooth boundary.  
Let $\gamma \in {\mathbb R}$ satisfy that $\Theta_0 < \gamma < 1$.
Then there exists $\epsilon_0 \in (0,1)$ such that, if
$\vec{A}=\vec{A}_B$ is a (curve of) vector field(s) satisfying \eqref{eq:UnifEst} with the given $C_0$,
and $U_B(x)$ is given by
\begin{align}
U_B(x) = \begin{cases}
\gamma B\;, & \text{ if } \dist(x, \partial\Omega) > 2 B^{-\frac{1}{8}} \\
\Theta_0 B - C_1 \tilde{k}(s) B^{\frac{1}{2}} - \epsilon_0^{-1}B^{\frac{1}{4}} \;, & \text{ if } \dist(x, \partial\Omega) \leq 2 B^{-\frac{1}{8}}\;,
\end{cases}
\end{align}
with $\tilde{k}(s) := k_{\rm max} - \epsilon_0 K(s)$, 
then
\begin{align}
\label{eq:Extra}
q_{B\vec{A}}[u] \geq \int_{\Omega} U_B(x) |u(x)|^2\,dx\;,
\end{align}
for all $u \in W^{1,2}(\Omega)$ and all $B>\epsilon_0^{-1}$.
\end{thm}

For comparison we include the following result from \cite[Proposition~3.7]{He-Pan}. 

\begin{thm}
\label{thm:UnifLowerBounds2}~\\
Let $C_0>0$ be given and let $\Omega$ be a bounded domain with smooth boundary. 
Then there exists $\epsilon_0> 0$ such that if
$\vec{A}=\vec{A}_B$ is a (curve of) vector field(s) satisfying \eqref{eq:UnifEst} with the given $C_0$,
and $U_B(x)$ is given by
\begin{align}
U_B(x) = \begin{cases}
(1-\epsilon_0^{-1} B^{-\frac{1}{6}}) B\;, & \text{ if } \dist(x, \partial\Omega) > 2 B^{-\frac{1}{6}} \\
\Theta_0 B - C_1 k(s) B^{\frac{1}{2}} - \epsilon_0^{-1}B^{\frac{1}{3}} \;, & \text{ if } \dist(x, \partial\Omega) \leq 2 B^{-\frac{1}{6}}\;.
\end{cases}
\end{align}
Then
\begin{align}
q_{B\vec{A}}[u] \geq \int_{\Omega} U_B(x) |u(x)|^2\,dx\;,
\end{align}
for all $u \in W^{1,2}(\Omega)$ and all $B>\epsilon_0^{-1}$.
\end{thm}

\begin{remark}~\\
The advantage of the result in Theorem~\ref{thm:UnifLowerBounds} compared to Theorem~\ref{thm:UnifLowerBounds2} is the improved error estimate ($B^{\frac{1}{4}}$ compared to $B^{\frac{1}{3}})$. However, the disadvantage is that the curvature $k(s)$ has been replaced by the smaller function $\tilde{k}(s)$.
\end{remark}

In the proof of Theorem~\ref{thm:UnifLowerBounds} we will use the following result from \cite{BPT}.

\begin{thm}
\label{lem:BPT}~\\
Let $\mu^{(1)}(h,b,D(0,R))$ be the ground state energy of the operator in \eqref{eq:MagQuad} in the case where $B=b$, $\vec{A} = \vec{F}$ (i.e. $\curl B\vec{A} = b$ is independent of $x \in \Omega$), $\Omega = D(0,R)$; the disc of radius $R$, and where $p_{B\vec{A}} = (-ih \nabla - B\vec{A})$. Then there exists $C>0$ such that,  if
$$
bR^2/h \geq C\;,
$$
then
\begin{align}
\mu^{(1)}(h,b,D(0,R)) \geq \Theta_0 b h - C_1 b^{1/2} h^{3/2}/R - C h^2 R^{-2}\;.
\end{align}
\end{thm}

\begin{remark}$\,$
\begin{itemize}
\item Clearly, in the case of a disc, the curvature $k$ is constant, $k = R^{-1}$.
\item The technical condition in Theorem~\ref{thm:UnifLowerBounds}---that $\Omega$ is bounded---is only imposed for the convenience of being able to apply Theorem~\ref{lem:BPT}. In the general case (for instance for exterior domains) one should instead follow and improve an analysis given in \cite[Section 10]{He-Mo}. However that would carry us too far astray here, so we only state and prove Theorem~\ref{thm:UnifLowerBounds} in the case of a bounded $\Omega$\,.
  \end{itemize}
\end{remark}

\begin{cor}
\label{cor:CorEnergyB1-4}~\\
Under the assumptions of Theorem~\ref{thm:UnifLowerBounds}, we have
\begin{align}
q_{B\vec{A}}[u] \geq \|u \|^2 \big(\Theta_0 B - C_1 k_{\rm max} B^{1/2} - {\mathcal O}(B^{1/4})\big)\;,
\end{align}
for all $u \in W^{1,2}(\Omega)$ and all $B$ sufficiently big. Here the ${\mathcal O}(B^{1/4})$ remainder only depends on $\Omega$ and on the constant $C_0$ in \eqref{eq:UnifEst}.
\end{cor}

\begin{proof}~\\
Corollary~\ref{cor:CorEnergyB1-4} clearly follows from Theorem~\ref{thm:UnifLowerBounds} upon estimating 
$$
U_B(x) \geq \inf_{y\in \Omega} U_B(y)\;.
$$
\end{proof}

\begin{proof}[Proof of Theorem~\ref{thm:UnifLowerBounds}]~\\
%
Let $u=u_{B,\vec{A},\epsilon_0}$ be a ground state of ${\mathcal H}_{B\vec{A}} - U_B$. Notice, that (for fixed $B, \epsilon_0$) it is clear that the spectrum of ${\mathcal H}_{B\vec{A}}  - U_B$ consists of a sequence of eigenvalues whose only accumulation point is $+\infty$. Therefore such a ground state $u$ exists.
Let $\lambda=\lambda_{B,\vec{A},\epsilon_0}$ be the associated ground state energy. We will prove that if $\epsilon_0$ is sufficiently small, then $\lambda\geq0$.

Notice that since $\vec{A}$ satisfies \eqref{eq:UnifEst}, we get, for all $u \in W^{1,2}_0(\Omega)$,
\begin{align}
q_{B\vec{A}}[u] \geq B \int |u(x)|^2  \,\curl \vec{A}\,dx \geq  B (1+ {\mathcal O}(B^{-\frac{1}{4}}))\;.
\end{align}
Therefore, $u$ satisfies the Normal Agmon estimates, Theorem~\ref{thm:agmon} (see \cite[(6.25) and (6.26)]{He-Mo}).

We can find a sequence $\{ s_{j,B} \}_{j=0}^{N(B)}$ in ${\mathbb R}/|\partial \Omega|$ and a partition of unity $\{ \tilde{\chi}_{j,B} \}_{j=0}^{N(B)}$ on ${\mathbb R}/|\partial \Omega|$ such that
$\supp \tilde{\chi}_{j,B} \cap \supp \tilde{\chi}_{k,B} = \emptyset$ if $j \notin \{k-1, k , k+ 1\}$ (with the convention that $N(B) + 1 = 0\,$, $-1 = N(B)$). Furthermore, we may impose the conditions, for some constant $C$ and for $B\geq B_0$ sufficiently big:
\begin{align}
\supp \tilde{\chi}_{j,B} &\subset s_{j,B}  + [-B^{-\frac{1}{8}}, B^{-\frac{1}{8}}]\;,  \nonumber\\
\sum_{j} \tilde{\chi}_{j,B}^2 &= 1 \;,
\quad\quad\quad
\sum_{j} | \nabla \tilde{\chi}_{j,B}|^2 \leq  C B^{\frac{1}{4}} \;.
\end{align}
We will always choose the $s_{j,B}$ such that $|s_{j,B}| \leq |\partial \Omega|/2\,$.

Let $\chi_1, \chi_2$ be a standard partition of unity on ${\mathbb R}$~:
\begin{align}\label{eq:chi12}
\chi_1^2 + \chi_2^2 &= 1\;, &
\supp \chi_1 &\subset (-2,2)\;, &
\chi_1 &= 1 \text{ on a nbhd of } [-1,1]\;.
\end{align}
Let us define
\begin{align}
\label{defthetajh}
\chi_{j,B}(s,t) &= \tilde{\chi}_{j,B}(s) \chi_1(B^{\frac{1}{8}}t)\;,\nonumber\\
\theta_{j,B} (x)&= \chi_j(B^{\frac{1}{8}}t(x))\;,\; \text{ for } j=1,\,2\;.
\end{align}
We will also consider $\chi_{j,B}$ as a function on $\Omega$ (by passing to boundary coordinates) without changing the notation.

Using the standard localization formula and the Normal Agmon estimates, we get
\begin{align*}
\lambda &\geq
\sum_{j} \langle \chi_{j,B} u  \, | \,\big( {\mathcal H} - U_B\big) \chi_{j,B} u \rangle
+ \int (B \,\curl \vec{A}-U_B) |\theta_{2,B} u|^2\,dx \\
&\quad\quad\quad\quad -
C\,B^{1/4} \int_{\{B^{-\frac{1}{8}} \leq t(x) \leq 2B^{-\frac{1}{8}}\}} |u|^2\,dx \\
& \geq 
\sum_{j} \langle \chi_{j,B} u  \, | \,\big( {\mathcal H} - U_B\big) \chi_{j,B} u \rangle + {\mathcal O}(B^{-\infty})\;.
\end{align*}
Modulo choosing $\epsilon_0$ sufficiently small, it therefore suffices to prove that
\begin{align}
\label{eq:suffices}
 \sum_{j} \Big\langle \chi_{j,B} u  \, | \, \big( {\mathcal H} - \Theta_0 B + 
C_1 \tilde{k}(s) B^{\frac{1}{2}} \big)\chi_{j,B} u \Big \rangle 
\geq - C B^{\frac{1}{4}} \int \sum_j |\chi_{j,B} u|^2\,dx\;.
\end{align}
Proving \eqref{eq:suffices} will finish the proof of Theorem~\ref{thm:UnifLowerBounds}.
We will write each of the terms in $\langle \cdot \,|\, \cdot \rangle$ in boundary coordinates and compare with the similar term with fixed curvature. 

\noindent{\bf Proof of \eqref{eq:suffices}.}\\
Since $\Omega$ is bounded we have $k_{\rm max}>0$. Outside a small neighborhood of the boundary points $s$ with $k(s) = k_{{\rm max}}$, the result of Theorem~\ref{thm:UnifLowerBounds2} is stronger than that of Theorem~\ref{thm:UnifLowerBounds}. 
Therefore, by Theorem~\ref{thm:UnifLowerBounds2} (for $B$ sufficiently large), if $j$ is such that $k$ takes negative values on $\supp \chi_{j,B}$, then
\begin{align}
\label{eq:sufficesNegative}
\Big\langle \chi_{j,B} u  \, | \, \big( {\mathcal H} - \Theta_0 B + 
C_1 \tilde{k}(s) B^{\frac{1}{2}} \big)\chi_{j,B} u \Big \rangle 
\geq - C B^{\frac{1}{4}} \int |\chi_{j,B} u|^2\,dx\;.
\end{align}
So only terms in \eqref{eq:suffices} with $k(s_{j,B}) \approx k_{{\rm max}}>0$ need to be analyzed finer than what Theorem~\ref{thm:UnifLowerBounds2} gives, and below we will therefore only consider $j$ such that $k(s_{j,B}) > 0$.

Using Lemma~\ref{lem:GoodGaugeImp}
we may assume that, on each of the sets $\supp \chi_{j,B}$, the gauge is chosen such that the expression for $\vec{A}$ in boundary coordinates is as in \eqref{eq:VeryGoodGauge}. In the estimates below, the constants $C$ will be uniform in $j$.
Define 
\begin{align*}
\tilde{A}_{1}(s,t) &= -t(1-tk(s)/2) + t^2 b(s,t)\;, &\;a(s,t)&=1-tk(s)\;.
\end{align*}
We know from \eqref {eq:UnifEst} and Lemma~\ref{lem:GoodGaugeImp} that 
\begin{align}
\| b \|_{L^{\infty}_{\rm loc}} \leq C B^{-\frac{1}{4}}\;.
\end{align}
Define 
\begin{equation}
{\mathcal B}_{j,B} := \int e[\chi_{j,B} u](s,t) \,ds\,dt \;, \label{defBjh}
\end{equation}
with
\begin{equation}
e[f] := a^{-1} |(D_s - B\tilde{A}_{1}) f |^2 + a |D_t f|^2\;.
\end{equation}
Similarly, we define 
\begin{align*}
k_{j,B} &= k(s_{j,B})\;, &
\;\tilde{A}_{1,j,B}(s,t)& = -t(1-t\frac{k_{j,B}}{2})\;,\\
k_{j,B}' &= k'(s_{j,B})\;, &
\;a_{j,B}&=1-tk_{j,B}\;,
\end{align*}
and
\begin{equation}
{\mathcal A}_{j,B} := \int e_{j,B}[\chi_{j,B} u](s,t) \,ds\,dt \;,\label{defAjh}\end{equation}
with
\begin{equation}
e_{j,B}[f] := a_{j,B}^{-1} |(D_s - B\tilde{A}_{1,j,B}) f |^2 + a_{j,B} |D_t f|^2\;.
\end{equation}
We observe that  ${\mathcal B}_{j,B} = \langle \chi_{j,B} u\, |\, {\mathcal H} \chi_{j,B} u \rangle$ and will compare ${\mathcal B}_{j,B}$
 and ${\mathcal A}_{j,B}$.
We clearly have
\begin{align}
\label{eq:ef}
e[\chi_{j,B} u](s,t) &= e_{j,B}[\chi_{j,B} u](s,t) + f_1(s,t) + f_2(s,t) - f_3(s,t)\;,\end{align}
with
\begin{align*}
f_1 &= (a^{-1}-a_{j,B}^{-1}) |(D_s - B\tilde{A}_{1}) \chi_{j,B} u |^2 
+ (a - a_{j,B}) |D_t (\chi_{j,B} u)|^2\;,\\
f_2 &= a_{j,B}^{-1} B^2 | (\tilde{A}_{1} - \tilde{A}_{1,j,B})  \chi_{j,B} u|^2\;,\\
f_3 &= 2 a_{j,B}^{-1} B \Re \Big\{(\tilde{A}_{1} - \tilde{A}_{1,j,B})  \chi_{j,B} \overline{u}
 (D_s - B\tilde{A}_{1,j,B}) \chi_{j,B} u \Big\}\;.
\end{align*}
Notice that for $s \in s_{j,B} + [-B^{-\frac{1}{8}}, B^{-\frac{1}{8}}]\,$, we have,
\begin{align}\label{eq:kappa}
|k(s) - k_{j,B}| 
&= |s - s_{j,B}| \cdot \Big| \int_0^1 k'((1-\ell) s_{j,B} + \ell s)\,d\ell \Big| \nonumber\\
&\leq C B^{-\frac{1}{8}}\big(| k'_{j,B}| + B^{-\frac{1}{8}}\big)\; .
\end{align}
Thus,
\begin{align}\label{eq:comp_a}
\begin{array}{l}
|a - a_{j,B}| = t |k(s) - k_{j,B}| \leq C B^{-\frac{1}{8}}\big(| k_{j,B}'| + B^{-\frac{1}{8}}\big) t\;, \\
|a^{-1} - a_{j,B}^{-1} | \leq C B^{-\frac{1}{8}} \big(| k_{j,B}'| + B^{-\frac{1}{8}}\big) t\;, \quad \text{ for } \quad t<2B^{-\frac{1}{8}}\;.
\end{array}
\end{align}
We estimate, using \eqref{eq:comp_a}, for any $\eta > 0\,$,
\begin{align}\label{eq:f1}
|f_1&(s,t)| \leq C B^{-\frac{1}{8}} \big(| k_{j,B}'| + B^{-\frac{1}{8}}\big) t \,e[\chi_{j,B}u](s,t)\nonumber \\&\leq
C' \eta B^{-\frac{1}{2}} \big(| k_{j,B}'| + B^{-\frac{1}{8}}\big)^2 \, e[\chi_{j,B}u](s,t) +
C' \eta^{-1} B^{\frac{1}{4}} t^2 \, e[\chi_{j,B}u](s,t) \;. 
\end{align}
We also  estimate $f_2$ and $f_3$ by
\begin{align}\label{eq:f2}
f_2(s,t) &\leq C B^{\frac{7}{4}} t^4 |\chi_{j,B} u|^2\;, 
\end{align}
and
\begin{align}\label{eq:f3}
|f_3(s,t)| &\leq 2 C t^2 B^{\frac{7}{8}} \big(| k_{j,B}'| + B^{-\frac{1}{8}}\big) a_{j,B}^{-1} \Big|\chi_{j,B} \overline{u}
 (D_s - B\tilde{A}_{1,j,h}) \chi_{j,B} u \Big| \nonumber\\
 &\leq
 C' \eta^{-1} B^{\frac{9}{4}} t^4 |\chi_{j,B} u|^2+
 C' \eta B^{-\frac{1}{2}} \big(| k_{j,B}'| + B^{-\frac{1}{8}}\big)^2 e_{j,B} [\chi_{j,B}u](s,t)\;. 
\end{align}
Thus, we get by combining \eqref{eq:ef} with \eqref{eq:f1}, \eqref{eq:f2}, and \eqref{eq:f3} and integrating,
\begin{align*}
\{1 + C \eta B^{-\frac{1}{2}}(|k_{j,B}'| + B^{-\frac{1}{8}})^2\} {\mathcal B}_{j,B} 
&\geq \{1 - C \eta B^{-\frac{1}{2}}(|k_{j,B}'| + B^{-\frac{1}{8}})^2\}{\mathcal A}_{j,B} \nonumber\\
&\quad- C \eta^{-1} B^{\frac{1}{4}} \int t^2 e[\chi_{j,B} u](s,t)\,ds\,dt \nonumber\\
&\quad- C B^{\frac{9}{4}}( B^{-\frac{1}{2}} +  \eta^{-1} ) \int t^4 |\chi_{j,B} u|^2\,ds\,dt\;.
\end{align*}
This gives (with a new constant $C$)
\begin{align}\label{eq:aB}
{\mathcal B}_{j,B} &\geq \{1 - C \eta B^{-\frac{1}{2}}(|k_{j,B}'|^2 + B^{-\frac{1}{4}})\}{\mathcal A}_{j,B} \nonumber\\
&\quad- C \eta^{-1} B^{\frac{1}{4}} \int t^2 e[\chi_{j,B} u](s,t)\,ds\,dt \nonumber\\
&\quad- C B^{\frac{9}{4}}( B^{-\frac{1}{2}} +  \eta^{-1} ) \int t^4 |\chi_{j,B} u|^2\,ds\,dt\;.
\end{align}
From Theorem~\ref{lem:BPT} we get the estimate
\begin{align}\label{eq:BPT}
{\mathcal A}_{j,B} &\geq \Big(\Theta_0 B - C_1 k_{j,B} B^{\frac{1}{2}} - C \Big) \| \chi_{j,B} u\|^2\;.
\end{align}
Therefore, with $K(s) := (k_{{\rm max}} - k(s))$, $K_{j,B} := (k_{{\rm max}} - k_{j,B})$, and making a Taylor expansion of $k(s)-k_{j,B}$, we get
\begin{align}\label{eq:pos2ndDerTilde}
&\Big\{1 - C \eta B^{-\frac{1}{2} }  \big(| k_{j,B}'|^2 + B^{-\frac{1}{4}}\big) \Big\}
{\mathcal A}_{j,B} \nonumber \\
&\quad\quad\quad\quad\quad\quad-
\Big(\Theta_0 B - C_1\{ k_{{\rm max}}-\epsilon_0K(s)\} B^{\frac{1}{2}}\Big) \| \chi_{j,B} u\|^2 \nonumber \\
&= C_1 \Big(K_{j,B} - \epsilon_0[ K_{j,B} + C' |k_{j,B}'| B^{-\frac{1}{8}}] - C' \eta \Theta_0 |k_{j,B}'|^2 \Big) B^{\frac{1}{2} }\| \chi_{j,B} u\|^2
\nonumber \\
&\quad\quad\quad\quad\quad\quad
+ {\mathcal O}(B^{\frac{1}{4} })\| \chi_{j,B} u\|^2 \;.
\end{align}
By definition, the function $K(s)$, which can also be identified with a periodic function on $\mathbb R$,  satisfies, for some $C>0$,
\begin{align}
\label{eq:elem}
K(s)\geq 0\;, \quad\quad
K''(s)\leq C\;.
\end{align}
Therefore, since $K'(s) = k'(s)$, we have the elementary inequality\footnote{Using \eqref{eq:elem} we find, for all $s,\sigma \in {\mathbb R}$,
$$
0\leq K(\sigma) \leq K(s) + K'(s)(\sigma-s) + C (\sigma-s)^2.
$$
Upon setting $\sigma= s-\frac{K'(s)}{2C}$ we get \eqref{eq:kprimecarre} with $C'=4C$.
},
\begin{align}
\label{eq:kprimecarre}
|k'(s)|^2 \leq C' K(s)\;,
\end{align}
for some constant $C'>0$ and all $s\in {\mathbb R}$.
So we see that for $\epsilon_0, \eta$ sufficiently small we get
\begin{align}\label{eq:pos2ndDer}
&\Big\{1 - C \eta B^{-\frac{1}{2} }  \big(| k_{j,B}'|^2 + B^{-\frac{1}{4}}\big) \Big\}
{\mathcal A}_{j,B}
 \nonumber \\
&\quad\quad\quad\quad\quad\quad-
\Big(\Theta_0 B - C_1\{ k_{{\rm max}}-\epsilon_0K(s)\} B^{\frac{1}{2}}\Big) \| \chi_{j,B} u\|^2 \nonumber \\
&\geq -C'  B^{\frac{1}{4}}  \| \chi_{j,B} u\|^2\;,
\end{align}
for some $C'>0$.
Therefore, we get from \eqref{eq:aB}, for the given choice of $\eta$,
\begin{align}\label{eq:AlmostThere}
\sum_j {\mathcal B}_{j,B} &
\geq \Big(\Theta_0 B - C_1 \tilde{k}(s) B^{\frac{1}{2}} - C B^{\frac{1}{4}}\Big) \| u\|^2 \\
&\quad - C B^{\frac{1}{4}} \int \sum_j  t^2 e[\chi_{j,B} u](s,t)\,ds\,dt 
- C B^{\frac{9}{4}} \int t^4 | u|^2\,ds\,dt\;.\nonumber
\end{align}
The normal Agmon estimates and easy manipulations (as in \cite{He-Mo})  give that the last two terms in \eqref{eq:AlmostThere} are bounded by $C B^{\frac{1}{4}}$. Therefore, \eqref{eq:suffices} follows from \eqref{eq:AlmostThere}.

This finishes the proof of Theorem~\ref{thm:UnifLowerBounds}.
\end{proof}

\subsection{Agmon estimates in the tangential direction}~\\
Theorem~\ref{thm:UnifLowerBounds} can be used to obtain exponential localization estimates in the tangential ($s$-)variable both for the linear and non-linear problems. These estimates are similar to Theorems~\ref{thm:agmon} and~\ref{thm:agmonNL} and are given in Theorems~\ref{thm:Agmons} and~\ref{thm:nucleation}.

\begin{thm}[Uniform Tangential Agmon estimates]
\label{thm:Agmons}~\\
Let $C_0 >0$ be given and let $\Omega$ be a bounded domain with smooth boundary.
Then there exist $C$, $\alpha>0$, such that if
$\vec{A} = (\vec{A}_B)_{B \in [C, +\infty)}$ is a family of vector fields satisfying \eqref{eq:UnifEst} (with the given $C_0$)
and
$(u_B)_{B\in[C,\infty[}$ is a family of  normalized eigenfunctions of ${\mathcal H}_{B\vec{A}}$ with corresponding eigenvalue $\lambda(B)$ satisfying the bound
\begin{align}\label{eq:EVbound}
\lambda = \lambda(B\vec{A}) \leq \Theta_0 B - C_1 k_{{\rm max}} B^{\frac{1}{2}} + C_0 B^{\frac{1}{4}}\;\;,\; \forall B \geq C\;,
\end{align} 
and if $\chi_1 \in C_0^{\infty}$ is the function from \eqref{eq:chi12}, then, for all $ B \geq C$, 
\begin{align}
\label{eq:Agmons}
\int_{\Omega} e^{2\alpha B^{\frac{1}{4}}K(s) } \chi_1^2(B^{\frac{1}{8}}t(x)) \Big\{
|u_B(x)|^2 + B^{-1}\big|(-i\nabla - B\vec{A}(x)) u_B(x)\big|^2 \Big\}\,dx \leq C\;.
\end{align}
\end{thm}

\begin{proof}~\\
The proof of Theorem~\ref{thm:Agmons} is similar to (but easier than) the proof of Theorem~\ref{thm:nucleation}, given in Section~\ref{sec:GenHC3} below, and will therefore be omitted.
\end{proof}

\begin{corollary}\label{CorNLT}~\\
  With the assumptions of the theorem and Assumption \ref{assump:omega}, for any $p \geq 2$, there exists a constant $C_p$ such that
\begin{equation}\label{2pT}
 ||\psi||_{L^2(\Omega)}
\leq C_p \kappa^{ - \frac{5(p-2)}{8p}} ||\psi||_{L^p(\Omega)} 
\;.
\end{equation}
\end{corollary}
This can also be extended without additional difficulties to the case
 when $K$ has isolated zeros of finite order. 

\subsection{An alternative approach to $\lambda_1(B\vec{A})$}~\\
In the case where $\lambda_1(B)$ is known to very high precision, it is advantageous to estimate $\lambda_1(B\vec{A})$ by first approximating by constant field and then using the knowledge of $\lambda_1(B)$. This is the result in Theorems~\ref{thm:unifSpec2} and~\ref{thm:unifSpecMod} below. 
Remember that $\lambda_1(B \vec{A})$ is the lowest eigenvalue (bottom of the spectrum) of ${\mathcal H}(B \vec{A})$.

\begin{thm}
\label{thm:unifSpec2}~\\
Let $C_0>0$ be given.
Suppose that $\Omega$ is a smooth, bounded, simply-connected domain in ${\mathbb R}^2$ and that $\Omega$ is not a disc.
Then there exists $B_0, \epsilon_0, C>0$ such that, for all $B\geq B_0$ and  if $\vec{A}$ satisfies 
\eqref{eq:UnifEst} with the given $C_0$,
then
\begin{align}
\label{eq:EstNoDisc}
\lambda_1(B \vec{A}) \geq \lambda_1(B) - C \big(\| \curl \vec{A} - 1 \|_{C^1(\Omega)} \sqrt{B} + e^{-\epsilon_0 \sqrt[4]{B}}\big)\;.
\end{align}
If $\Omega$ is a disc, then \eqref{eq:EstNoDisc} is replaced by
\begin{align}
\lambda_1(B \vec{A}) \geq \lambda_1(B) - C \big(\| \curl \vec{A} - 1 \|_{C^1(\Omega)} \sqrt{B} + 1\big)\;.
\end{align}
\end{thm}

\begin{proof}~\\
We consider first the case where $\Omega$ is not a disc.

Let $\phi_{B\vec{A}}$ be a normalized ground state of ${\mathcal H}(B \vec{A})$.
Since, by Corollary~\ref{cor:CorEnergyB1-4}, $\lambda_1(B) = \Theta_0 B  - C_1 k_{{\rm max}} B^{\frac{1}{2}} + {\mathcal O}(B^{1/4})$, we may assume that 
\begin{align*}
\lambda_1(B\vec{A})  \leq \Theta_0 B - C_1 k_{{\rm max}} B^{\frac{1}{2}} + C_0 B^{\frac{1}{4}}\;,
\end{align*} 
(if not, there is nothing to prove).
Then the normal Agmon estimates (given in Theorem~\ref{thm:agmon}) and the tangential Agmon estimates
(given in Theorem~\ref{thm:Agmons}) give exponential localization estimates on $\phi_{B\vec{A}}$.
Since $\Omega$ is not a disc, there exists a $\sigma_0 \in {\mathbb R}/|\partial \Omega|$ such that 
$$
k(\sigma_0) = k_{\rm min} = \min_{s \in {\mathbb R}/|\partial \Omega|} k(s) \neq k_{{\rm max}}\;.
$$
Choose $\epsilon>0$ such that $k(s) \leq \frac{k_{\rm min} + k_{\rm max}}{2}$ on $|s-\sigma_0| \leq \epsilon$.
Let $f_1^2 + f_2^2 = 1$ be a partition of unity on $\Omega$ such that 
\begin{align*}
&f_1 = 1 \text{ on } \{ t \leq t_0/2 \} \cap \{| s - \sigma_0| \geq \epsilon\},\\ 
&\supp f_1 \subseteq \{ t \leq t_0 \} \cap\{| s - \sigma_0| \geq \epsilon/2\}\;.
\end{align*}
By the standard localization formula,
\begin{align*}
\lambda_1(B\vec{A}) = q_{B\vec{A}}[\phi_{B\vec{A}}] &= 
q_{B\vec{A}}[f_1 \phi_{B\vec{A}}] + q_{B\vec{A}}[f_2 \phi_{B\vec{A}}] \\
&\quad\quad
-  \int_{\Omega}\big( |\nabla f_1|^2 +  |\nabla f_2|^2\big) |\phi_{B\vec{A}}|^2\;dx\;.
\end{align*}
Using the Agmon estimates, we therefore find for some $C, \epsilon_0$,
\begin{align}
\label{eq:Peerlim}
\lambda_1(B\vec{A}) \geq 
q_{B\vec{A}}[f_1 \phi_{B\vec{A}}] -C e^{-\epsilon_0 \sqrt[4]{B}}\;.
\end{align}
Using Lemma~\ref{lem:GoodGaugeImp} (in the situation given by \eqref{eq:VeryGoodGauge}) we know that we can choose a gauge $\varphi$ on $\supp f_1$ such that
$$
\tilde{A} - \tilde{F} - \nabla \varphi = t^2 \epsilon(s,t)
$$
where $\tilde{A}, \tilde{F}$ are $\vec{A}$ and $\vec{F}$ transformed to boundary coordinates and where
$\| \epsilon(s,t) \|_{L^{\infty}}$ is controlled by $\| \curl \vec{A} - 1 \|_{C^1}$.
Therefore, we can estimate, for all $\rho > 0$,
\begin{align}
\label{eq:aellebaelle}
q_{B\vec{A}}[f_1 \phi_{B\vec{A}}] \geq
(1-\rho) q_{B\vec{F}}[f_1 e^{iB \varphi} \phi_{B\vec{A}}]-
\rho^{-1}B^2  \int \big| t^2 \epsilon(s,t)  f_1 \phi_{B\vec{A}}\big |^2\;dx\;.
\end{align}
Using the normal Agmon estimates, we therefore get from \eqref{eq:Peerlim} and \eqref{eq:aellebaelle}
\begin{align}
\lambda_1(B\vec{A}) \geq 
(1-\rho) \lambda_1(B) \| f_1 \phi_{B\vec{A}}\|^2 
-C \rho^{-1} B \| \curl \vec{A} - 1 \|_{C^1(\overline{\Omega})}^2
-C e^{-\epsilon_0 \sqrt[4]{B}}\;.
\end{align}
We finish the proof of \eqref{eq:EstNoDisc} by applying the normal and tangential Agmon estimates again (to remove the localization $f_1$) and 
by choosing $\rho=\| \curl \vec{A} - 1 \|_{C^1(\overline{\Omega})}/\sqrt{B}$ (using that $\lambda_1(B) \leq B$ for large $B$).

When $\Omega$ is a disc, we make the partition of unity $f_1, f_2$ as follows.
\begin{align*}
&f_1 = 1 \text{ on } \{ t \leq t_0/2 \} \cap \{| s | \leq \tfrac{|\partial \Omega|}{4}\},\\ 
&\supp f_1 \subseteq \{ t \leq t_0 \} \cap\{| s | \leq \tfrac{3|\partial \Omega|}{4}\}\;.
\end{align*}
The localization error $\int_{\Omega}\big( |\nabla f_1|^2 +  |\nabla f_2|^2\big) |\phi_{B\vec{A}}|^2\;dx$ then becomes of unit size, and we get
\begin{align}
\lambda_1(B\vec{A}) \geq  
q_{B\vec{A}}[f_1 \phi_{B\vec{A}}] + q_{B\vec{A}}[f_2 \phi_{B\vec{A}}] + {\mathcal O}(1)\;.
\end{align}
Both $q_{B\vec{A}}[f_1 \phi_{B\vec{A}}]$ and $q_{B\vec{A}}[f_2 \phi_{B\vec{A}}]$ are now estimated as above. This finishes the proof.
\end{proof}

When $\| \curl \vec{A} - 1 \|_{C^1(\Omega)}$ is very small, the exponentially small error in Theorem~\ref{thm:unifSpec2} is too expensive. Therefore we need the following simpler result.

\begin{thm}
\label{thm:unifSpecMod}~\\
Suppose $\Omega$ is a smooth, bounded, simply-connected domain in ${\mathbb R}^2$.
Then there exist $B_0 > 0$ and, for all $p>2$, $C_p >0$ such that
\begin{align}
\label{eq:e1}
\lambda_1(B \vec{A}) \geq \lambda_1(B) - C_p B^{3/2} \| \curl \vec{A} - 1 \|_{L^p(\Omega)} \;,
\end{align}
when $B\geq B_0$.
\end{thm}

\begin{proof}~\\
To prove the estimate \eqref{eq:e1}, we write $b=\curl \vec{A}-1 \in C_0^1(\Omega)$ and define
\begin{align*}
\Phi(x) &= \frac{1}{2\pi} \int_{\Omega} (\log |x-y| ) b(y)\,dy\;, &
\vec{a} &= (-\partial_2 \Phi, \partial_1 \Phi)\;.
\end{align*}
Then $\curl \vec{a} = \Delta \Phi = b$ in $\Omega$, and for all $p>2$,
\begin{align}
\label{eq:ainfty}
\| \vec{a} \|_{L^{\infty}} \leq C_p \| b \|_{L^{p}(\Omega)}\;.
\end{align}
Since $\Omega$ is simply connected there exists a choice of gauge $\varphi$ such that $\vec{A} - \vec{F} - \nabla \varphi= \vec{a}$.
With $\bar{\phi}_B = e^{-iB\varphi} \phi_{B\vec{A}}$, we get for all $0<\rho$,
\begin{align*}
\lambda_1(B \vec{A}) &= 
\int_{\Omega} \big| \big(-i\nabla - B (\vec{A}- \nabla \varphi)\big)\bar{\phi}_{B}\big|^2\,dx \\
&\geq
(1-\rho) \int_{\Omega} | (-i\nabla - B \vec{F})\bar{\phi}_{B}|^2\,dx
- \rho^{-1} B^2 \int_{\Omega} | \vec{a} \, \bar{\phi}_{B}|^2\,dx 
\end{align*}
This implies (notice that \eqref{eq:rhorow} is trivial for $\rho >1$) by \eqref{eq:ainfty}, for all $\rho > 0$,
\begin{align}
\label{eq:rhorow}
\lambda_1(B \vec{A}) 
&\geq (1-\rho) \lambda_1(B) -C \rho^{-1} B^2 \| b \|_{L^{p}(\Omega)}^2\;.
\end{align}
By choosing $\rho = \| b \|_{L^{p}(\Omega)} \sqrt{B}$, we get the desired estimate (using that $\lambda_1(B) \leq B$ for large $B$).
\end{proof}

\section{Proofs of Theorem~\ref{thm:GeneralHG3} and Theorem~\ref{thm:nucleation}}
\label{sec:GenHC3}
Let us start by recalling the following result from \cite[Prop.4.2]{He-Pan}.

\begin{thm}
\label{thm:elliptic}~\\
Suppose $(\psi, \vec{A})=(\psi_{\kappa, H}, \vec{A}_{\kappa, H})$ is a sequence of minimizers for the GL-functional. 
Suppose that the parameters $\kappa, H$ satisfy,
\begin{align}
\label{eq:AlmostHC3}
\big( \Theta_0^{-1} + g(\kappa) \big) \kappa \leq H < H_{C_3}(\kappa)\;,
\end{align}
for some function $g: {\mathbb R} \rightarrow {\mathbb R}$ with $\lim_{\kappa \rightarrow \infty} g(\kappa) = 0$.
Then there exists a constant $C>0$, such that for all $\kappa > C$, we have
\begin{align}
\| \curl \vec{A} - 1 \|_{C^1(\overline{\Omega})} &\leq \frac{C}{\sqrt{\kappa H}} \| \psi \|_{L^{\infty}(\Omega)}^2 \;, \\
\| \curl \vec{A} - 1 \|_{C^2(\overline{\Omega})} &\leq C \|\psi \|_{L^{\infty}(\Omega)}^2\;.
\end{align}
In particular,
\begin{align}
\| \curl \vec{A} - 1 \|_{C^1(\overline{\Omega})} &\leq C (\kappa H)^{-\frac{1}{2}}\;, &
\| \curl \vec{A} - 1 \|_{C^2(\overline{\Omega})} &\leq C\;.
\end{align}
\end{thm}

\begin{remark}~\\
In particular, Theorem~\ref{thm:elliptic} implies that the minimizing vector potential $\vec{A}_{\kappa, H}$ satisfies \eqref{eq:UnifEst} for $B = \kappa H$. Thus, we can apply the results from subsection~\ref{sec:UnifEnergy} on the magnetic quadratic form $q_{\kappa H \vec{A}_{\kappa, H}}$ appearing in the Ginzburg-Landau functional.
\end{remark}

\begin{proof}[Proof of Theorem~\ref{thm:GeneralHG3}]~\\
We consider first the case of the disc. 
Using Theorem~\ref{thm:disc}, we see that there exists $C>0$, such that if $\kappa >C$, and 
$$
H < \frac{\kappa}{\Theta_0} + \frac{C_1}{\Theta_0^{\frac{3}{2}}} k_{{\rm max}} - C/\kappa\;,
$$
then $\lambda_1(\kappa H) < \kappa^2$. Therefore, Theorem~\ref{thm:LetHalvdel} implies that
\begin{align}
\underline{H_{C_3}(\kappa)} \geq \underline{H_{C_3}^{\rm loc}(\kappa)} > \frac{\kappa}{\Theta_0} + \frac{C_1}{\Theta_0^{\frac{3}{2}}} k_{{\rm max}} - C/\kappa\;.
\end{align}
On the other hand, 
Theorem~\ref{thm:unifSpec2}, combined with Theorems~\ref{thm:disc} and~\ref{thm:elliptic}, implies the existence of $C'>0$, such that if $\kappa >C'$, $H$ satisfies
$$
\overline{H_{C_3}(\kappa)} \geq H > \frac{\kappa}{\Theta_0} + \frac{C_1}{\Theta_0^{\frac{3}{2}}} k_{{\rm max}} + C'/\kappa\;,
$$
and $(\psi, \vec{A})$ is a Ginzburg-Landau minimizer with parameters $(\kappa, H)$,
then necessarily $\lambda_1(\kappa H \vec{A}) > \kappa^2$.
But then $(\psi, \vec{A}) = (0,\vec{F})$.
This finishes the proof in the case of the disc.

Consider now the case of general $\Omega$. The lower bound in Corollary~\ref{cor:CorEnergyB1-4} combined with the matching upper bound from \cite{He-Mo} gives that
$$
\lambda_1(\kappa H) =
\Theta_0 \kappa H - C_1 k_{\rm max} (\kappa H)^{1/2} + {\mathcal O}((\kappa H)^{1/4}).
$$
From this it is an elementary calculation to prove that there exists $C>0$ large enough such that 
\begin{align*}
H &\leq \frac{\kappa}{\Theta_0} + \frac{C_1}{\Theta_0^{\frac{3}{2}}} k_{{\rm max}} - C \kappa^{-\frac{1}{2}}
\end{align*}
implies that
\begin{align*}
\lambda_1(\kappa H) < \kappa^2 - \sqrt{\kappa}\;,
\end{align*}
for all $\kappa >C$. 
In particular, $\lambda_1(\kappa H) < \kappa^2 - \sqrt{\kappa}$, for all $\kappa >C$, and
therefore Theorem~\ref{thm:LetHalvdel} implies that
\begin{align}
\underline{H_{C_3}} \geq \underline{H_{C_3}^{\rm loc}}\geq \frac{\kappa}{\Theta_0} + \frac{C_1}{\Theta_0^{\frac{3}{2}}} k_{{\rm max}} - C \kappa^{-\frac{1}{2}}\;.
\end{align}
To prove the opposite inequality, let $(\kappa, H)_n$ be a sequence of parameters (we will suppress the $n$ from the notation for convenience) such that
\begin{itemize}
\item $H\leq \overline{H_{C_3}(\kappa)}$,
\item there exists a non-trivial GL-minimizer $(\psi, \vec{A})$ when $(\kappa, H)$ is in the sequence.
\item $H\geq \frac{\kappa}{\Theta_0} + \frac{C_1}{\Theta_0^{\frac{3}{2}}} k_{{\rm max}} + C \kappa^{-\frac{1}{2}}$, where $C$ is a (big) constant to be chosen below. 
\end{itemize}
Then Theorem~\ref{thm:elliptic} and Corollary~\ref{cor:CorEnergyB1-4}  imply that, if the constant $C$ is chosen sufficiently big,
$$
\lambda_1(\kappa H \vec{A}) \geq \kappa^2 + \sqrt{\kappa}>\kappa^2\;.
$$
But this contradicts the non-triviality of the GL-minimizers and therefore proves that
\begin{align}
\overline{H_{C_3}}\leq \frac{\kappa}{\Theta_0} + \frac{C_1}{\Theta_0^{\frac{3}{2}}} k_{{\rm max}} + C \kappa^{-\frac{1}{2}}\;.
\end{align}
\end{proof}

We will now estimate the tangential size of the superconducting boundary layer when $H$ is very close to---but below---$H_{C_3}$.

\begin{proof}[Proof of Theorem~\ref{thm:nucleation}]~\\
The proof is a variant of the proof of a similar result obtained in \cite{He-Pan}.
Since $0 \leq H_{C_3}(\kappa) - H = \rho$, we get from Theorem~\ref{thm:GeneralHG3} that 
$$
H =  \frac{\kappa}{\Theta_0} + \frac{C_1}{\Theta_0^{\frac{3}{2}}} k_{{\rm max}} + {\mathcal O}(\kappa^{-\frac{1}{2}}) + \rho\;.
$$
But then an elementary calculation gives that
\begin{align}
\label{energyGap}
\kappa^2 - [ \Theta_0 \kappa H - C_1 k_{\rm max} (\kappa H)^{1/2} - {\mathcal O}((\kappa H)^{1/4})] = {\mathcal O}(\kappa \rho) + {\mathcal O}(\sqrt{\kappa})\;.
\end{align}

With $\chi_1, \chi_2$ being a standard partition of unity and using the equation \eqref{eq:GL} satisfied by the GL-minimizer $(\psi, \vec{A})$, we can calculate
\begin{align}
\label{eq:GLquad}
 \kappa^2 \big\| \chi_1(\kappa^{\frac{1}{4}} t) & \exp(\alpha \kappa^{\frac{1}{2}} K(s)) \psi \big\|^2 
 \nonumber\\
\geq &
\big \langle \chi_1^2(\kappa^{\frac{1}{4}} t)  \exp(2\alpha \kappa^{\frac{1}{2}} K(s)) \psi \, \big | \, {\mathcal H}(\kappa H \vec{A}) \psi \big \rangle
\nonumber \\
= &
q_{\kappa H \vec{A}} \big[ \chi_1(\kappa^{\frac{1}{4}} t)  \exp(\alpha \kappa^{\frac{1}{2}} K(s)) \psi\big] \nonumber \\
&\quad\quad\quad\quad\quad
-
\int_{\Omega} \big|\nabla \big(\chi_1(\kappa^{\frac{1}{4}} t)  \exp(\alpha \kappa^{\frac{1}{2}} K(s))\big) \big|^2 |\psi|^2 \,dx\;.
\end{align}
We can estimate the localization error as
\begin{align}
\int_{\Omega} \Big|\nabla \Big(\chi_1(\kappa^{\frac{1}{4}} t) & \exp(\alpha \kappa^{\frac{1}{2}} K(s))\Big) \Big|^2 |\psi|^2 \,dx \leq L_1 + L_2\;,
\end{align}
where
\begin{align*}
L_1 &:= C \alpha^2 \kappa  \int_{\Omega} |K'(s)|^2 \chi_1^2(\kappa^{\frac{1}{4}} t) \exp(2\alpha \kappa^{\frac{1}{2}} K(s)) |\psi|^2 \,dx\;,  \\
L_2 &:=
C\kappa^{\frac{1}{2}} \int_{\Omega}| \chi_1'(\kappa^{\frac{1}{4}} t)|^2  \exp(2\alpha \kappa^{\frac{1}{2}} K(s)) |\psi|^2 \,dx  \;.
\end{align*}
Using the Agmon estimates in the normal direction, Theorem~\ref{thm:agmonNL}, we get 
\begin{align}
\label{eq:L2Agmon}
  L_2 = \| \psi \|^2 {\mathcal O}(\kappa^{-\infty})\;.
\end{align}
Now, using Theorems~\ref{thm:UnifLowerBounds} and~\ref{thm:elliptic}, we have
\begin{align}
\label{eq:QuadBelow}
  q_{\kappa H \vec{A}} [ \chi_1(\kappa^{\frac{1}{4}} t) & \exp(\alpha \kappa^{\frac{1}{2}} K(s)) \psi] \nonumber \\
\geq &
\int \Big[ \Theta_0 \kappa H - C_1 k_{\rm max} (\kappa H)^{\frac{1}{2}} +
\epsilon_0 K(s)  (\kappa H)^{\frac{1}{2}} - {\mathcal O}((\kappa H)^{\frac{1}{4}})
\Big] \nonumber \\
&\quad\quad\times
\Big| \chi_1(\kappa^{\frac{1}{4}} t)  \exp(\alpha \kappa^{\frac{1}{2}} K(s)) \psi\Big|^2\,dx\;.
\end{align}
Using \eqref{energyGap}, \eqref{eq:GLquad}, and \eqref{eq:QuadBelow}, we therefore get, remembering that by \eqref{eq:kprimecarre}, we have $|K'(s)|^2 \leq C K(s)$,
\begin{align}
\label{eq:A}
  L_2 \geq &
\int \Big[ \epsilon_0 K(s)  (\kappa H)^{1/2} 
- C \alpha^2 \kappa |K'(s)|^2 +
{\mathcal O}(\sqrt{\kappa})+ {\mathcal O}(\kappa \rho)
\Big] \nonumber \\
&\quad\quad\quad\times
\Big| \chi_1(\kappa^{\frac{1}{4}} t)  \exp(\alpha \kappa^{\frac{1}{2}} K(s)) \psi\Big|^2\,dx\; \nonumber\\
\geq &
\int \Big[ \epsilon_0 (1-\alpha^2 C')K(s)  (\kappa H)^{1/2} +
{\mathcal O}(\sqrt{\kappa})+ {\mathcal O}(\kappa \rho)
\Big] \nonumber \\
&\quad\quad\quad\times
\Big| \chi_1(\kappa^{\frac{1}{4}} t)  \exp(\alpha \kappa^{\frac{1}{2}} K(s)) \psi\Big|^2\,dx\;.
\end{align}
We split the integral on the right hand side in \eqref{eq:A} as 
$$
\int_{\Omega} = \int_{\{ K(s) \geq S[\rho + \kappa^{-\frac{1}{2}}]\}} + 
\int_{\{ K(s) <  S[\rho + \kappa^{-\frac{1}{2}}]\}} =: {\mathfrak I}_1 + {\mathfrak I}_2\;,
$$
for some $S >0$. We will choose $S$ sufficiently large below. \\
By definition of ${\mathfrak I}_2$, we get
\begin{align}
\label{eq:I-2}
 | {\mathfrak I}_2|  \leq C(\kappa \rho + \sqrt{\kappa}) e^{2\alpha S(1 +\sqrt{\kappa} \rho)} \| \psi \|^2\;,
\end{align}
If $\alpha$ is sufficiently small and $S$ is sufficiently big, we have
\begin{align}
\label{eq:I-1}
{\mathfrak I}_1 \geq  (\sqrt{\kappa}+\kappa\rho) \int_{\{ K(s) \geq S[\rho + \kappa^{-\frac{1}{2}}]\}}
 \chi_1^2(\kappa^{\frac{1}{4}} t)  \exp(2\alpha \kappa^{\frac{1}{2}} K(s)) |\psi(x)|^2\,dx\;.
\end{align}
Combining the estimates \eqref{eq:A}, \eqref{eq:I-2}, \eqref{eq:I-1}, and \eqref{eq:L2Agmon}, we find
(for $\alpha$ sufficiently small and $S$ sufficiently big)
\begin{align}
\label{eq:Large}
 \int_{\{ K(s) \geq S[\rho + \kappa^{-\frac{1}{2}}]\}}
 \chi_1^2(\kappa^{\frac{1}{4}} t)  \exp(2\alpha \kappa^{\frac{1}{2}} K(s)) |\psi(x)|^2\,dx
\leq
C e^{C \sqrt{\kappa} \rho} \| \psi \|^2\;.
\end{align}
Evidently,
\begin{align}
\label{eq:small}
 \int_{\{ K(s) \leq S[\rho + \kappa^{-\frac{1}{2}}]\}}
 \chi_1^2(\kappa^{\frac{1}{4}} t)  \exp(2\alpha \kappa^{\frac{1}{2}} K(s)) |\psi(x)|^2\,dx
\leq
C e^{C \sqrt{\kappa} \rho} \| \psi \|^2\;.
\end{align}
Combining \eqref{eq:Large} and \eqref{eq:small} finishes the proof of the theorem.
\end{proof}

\section{An improved estimate on $\| \psi \|_{L^{\infty}}$}~\\
From the maximum principle, one gets that minimizers $(\psi, \vec{A})$ of the Ginzburg-Landau functional \eqref{eq:GL_F} satisfy
the estimate $\| \psi \|_{L^{\infty}} \leq 1$ independently of the values of $\kappa, H$. When $H$ is far below $H_{C_3}$, that is a very useful estimate, but in the region near $H_{C_3}$, this estimate is far from optimal and it is interesting to have a better control of $||\psi||_{\infty}$.\\
Let us start by a non-rigorous argument which should give a limit on what one can hope to prove.~\\
Multiplying the GL-equation \eqref{eq:GL} by $\overline{\psi}$ and integrating, one gets
\begin{align}
\label{eq:GLint}
Q_{\kappa H \vec{A}}(\psi) - \kappa^2 \| \psi \|_2^2 + \kappa^2 \| \psi \|_4^4 = 0\;.
\end{align}
Let $\delta = \kappa^2 - \lambda_1(\kappa H \vec{A}) $ be the spectral distance, then one expects that
\begin{align}
\label{eq:WRONG}
Q_{\kappa H \vec{A}}(\psi) - \kappa^2 \| \psi \|_2^2 &\approx - \delta \| \psi \|_2^2\;,&
\frac{\| \psi \|_4^4}{\| \psi \|_2^2} \approx \| \psi \|_{\infty}^2\;.
\end{align}
Therefore the GL-equation \eqref{eq:GLint} implies that
$$
\| \psi \|_{\infty} \approx C \sqrt{\delta} \kappa^{-1}\;.
$$
Unfortunately, it is difficult to justify the second estimate in \eqref{eq:WRONG} rigorously, so we will only obtain a  somewhat less accurate estimate.

\begin{proposition}
\label{lem:Linfty2}~\\
Let $\Omega$ be a bounded, simply-connected domain with smooth boundary. 
Suppose that $(\psi,\vec{A})_{\kappa,H}$ is a family of minimizers of the GL-functional with
$0<\rho = H_{C_3}(\kappa) - H = o(\kappa)$, as $\kappa \rightarrow \infty$.
For all $\epsilon >0$,
\begin{align}
\label{eq:InftyGen}
\| \psi \|_{L^{\infty}(\Omega)} \leq C \delta^{1/2} \kappa^{-\frac{1}{2} + \epsilon}\;.
\end{align}
If $\Omega$ satisfies Assumption~\ref{assump:GenNondegen}, and $\rho = {\mathcal O}(\kappa^{-\frac{1}{2}})$, then \eqref{eq:InftyGen} is improved to
\begin{align}
\label{eq:InftySpec}
\| \psi \|_{L^{\infty}(\Omega)} \leq C \delta^{1/2} \kappa^{-\frac{5}{8} + \epsilon}\;.
\end{align}
\end{proposition}

The proposition will be a consequence of the 
\begin{lemma}\label{lem:integral}~\\
Under the assumptions of the proposition,  for all $\epsilon_1, \epsilon_2 >0$ such that $\epsilon_1\leq 1+\epsilon_2$, there  exists a constant $C>0$, such that
if $\delta$ is the corresponding spectral distance, 
$0<\delta =  \kappa^2 - \lambda_1(\kappa H \vec{A})$, then
\begin{align}
\label{eq:InftyBasic}
\lambda \leq C \delta^{1/2} \kappa^{\epsilon_1 + \epsilon_2} \mu^{1+\epsilon_1}\;,
\end{align}
where  $\lambda = \| \psi \|_{\infty}$ and
 $\mu$ is defined by $\lambda \mu = \|\psi \|_2$.\end{lemma}

\begin{proof}[Proof of Lemma~\ref{lem:integral}]~\\
Before we start the real proof, we state the basic inequalities that we will use.
The estimates
\begin{align}
\label{eq:GLcons}
Q_{\kappa H \vec{A}}(\psi) &\leq \kappa^2 \|\psi \|_2^2 \;,&
\| \psi \|_4^4 \leq \frac{\delta}{\kappa^2} \| \psi \|_2^2\;,
\end{align}
are easy consequences of \eqref{eq:GLint}.
Furthermore,
from \cite[Prop.~4.2]{He-Pan}, we get the inequality
\begin{align}
\label{eq:HP}
\| \nabla_{\kappa H \vec{A}} \psi \|_{\infty} \leq C \sqrt{\kappa H} \| \psi \|_{\infty}\;.
\end{align}
By the Sobolev inequality and interpolation, we get that for all $ps > 2$, $0<s\leq 1$,
$$
\lambda \leq C \| |\psi| \|_{W^{s,p}} \leq C \| \psi \|_p^{1-s} \|
\nabla |\psi| \|_p^{s} + C \|\psi \|_p \;.
$$
We then use the diamagnetic and H\"older's inequalities on the right hand side~:
$$
\lambda\leq C \| \psi \|_p^{1-s}
\; \| (-i\nabla-\kappa H \vec{A}) \psi \|_{p}^{s}
 + C  \lambda
\mu^\frac 2p\;.
$$
Using \eqref{2p}, with $p=\infty$, we get that
\begin{equation}\label{majmu}
\mu \leq C \kappa^{-\frac 12}\;.
\end{equation}
So for $\kappa$ large enough, we obtain
$$
\lambda\leq C \| \psi \|_p^{1-s}
\; \| (-i\nabla-\kappa H \vec{A}) \psi \|_{p}^{s} \;.
$$
We now apply H\"older's inequality for each term of the right hand side~:
$$
\lambda\leq C \big\{ \lambda^{p-4} \| \psi \|_4^4 \big\}^{\frac{1-s}{p}}
\Big\{ \| (-i\nabla-\kappa H \vec{A}) \psi \|_{\infty}^{p-2} \|
(-i\nabla-\kappa H \vec{A}) \psi \|_2^2 \Big\}^\frac{s}{p}\;.
$$
We use \eqref{eq:GLcons} and \eqref{eq:HP} to get
\begin{align}
\lambda \leq C \big( \lambda^{p-2} \mu^2 \frac{\delta}{\kappa^2} \big)^{\frac{1-s}{p}}
( \lambda^p \kappa^p \mu^2 )^{\frac{s}{p}} =
C \lambda^{1 -2\frac{1-s}{p}} \delta^{\frac{1-s}{p}}\mu^{\frac{2}{p}}
\kappa^{s - 2 \frac{1-s}{p}}\;. 
\end{align}
This implies that
\begin{align}
\label{eq:beforeAgmon}
\lambda \leq C \delta^{\frac{1}{2}} \mu^{\frac{1}{1-s}} \kappa^{\frac{ps}{2(1-s)} -1}\;.
\end{align}
Write $\frac{1}{1-s} = 1 + \epsilon_1$ and $ps=2+\frac{2\epsilon_2}{1+\epsilon_1}$. Then we find \eqref{eq:InftyBasic}.
\end{proof}
~\\
\begin{proof}[Proof of Proposition~\ref{lem:Linfty2}]~\\
Applying \eqref{majmu} to \eqref{eq:InftyBasic}, with $\epsilon_1=\epsilon$, $\epsilon_2 = \frac{\epsilon}{2}$, we find \eqref{eq:InftyGen}.
If $\rho = {\mathcal O}(\kappa^{-\frac{1}{2}})$ and
under Assumption~\ref{assump:GenNondegen}, we can also apply the tangential Agmon estimates, Theorem~\ref{thm:nucleation} and its corollary \ref{CorNLT} with $p=4$, and therefore bound
\begin{align}
\mu\leq C \kappa^{-\frac{5}{8}}\;.
\end{align}
This implies \eqref{eq:InftySpec}.
\end{proof}

\begin{remark}~\\
Using that $0 \leq H_{C_3}(\kappa) - H = \rho$, we get from Theorem~\ref{thm:GeneralHG3} that 
$$
H =  \frac{\kappa}{\Theta_0} + \frac{C_1}{\Theta_0^{\frac{3}{2}}} k_{{\rm max}} + {\mathcal O}(\kappa^{-\frac{1}{2}}) + \rho\;.
$$
Therefore Corollary~\ref{cor:CorEnergyB1-4} and an elementary calculation gives
\begin{align}
\label{energyGap2}
\delta = \kappa^2 - [ \Theta_0 \kappa H - C_1 k_{\rm max} (\kappa H)^{1/2} - {\mathcal O}((\kappa H)^{1/4})] = {\mathcal O}(\kappa \rho) + {\mathcal O}(\sqrt{\kappa})\;.
\end{align}
Therefore, \eqref{eq:InftyGen} expressed in terms of $\rho$ becomes
\begin{align}
\| \psi \|_{L^{\infty}(\Omega)} \leq C ( \sqrt{\rho} \kappa^{\epsilon} + \kappa^{-\frac{1}{4}+\epsilon})\;.
\end{align}
Under Assumption~\ref{assump:GenNondegen}, and with $\rho = {\mathcal O}(\kappa^{-\frac{1}{2}})$,  \eqref{eq:InftySpec} becomes
\begin{align}
\| \psi \|_{L^{\infty}(\Omega)} \leq C ( \sqrt{\rho} \kappa^{-\frac{1}{8}+\epsilon} + \kappa^{-\frac{1}{4}+\epsilon})\;.
\end{align}
\end{remark}

\section{$\overline{H_{C_3}(\kappa)} = \overline{H_{C_3}^{\rm loc}(\kappa)}$ for all domains.}
\label{equal}
In this section we will prove Theorem \ref{thm:Identical} 
 stating that the local and global upper-fields are equal---with no extra hypothesis on the domain $\Omega$.\\

We know that $\overline{H_{C_3}(\kappa)} \geq \overline{H_{C_3}^{\rm loc}(\kappa)}$, so we only need to prove the opposite inequality.
Let $H\leq\overline{H_{C_3}(\kappa)}$. We may assume that $H>\overline{H_{C_3}^{\rm loc}(\kappa)}$, so that
\begin{align}
\label{eq:Contradiction}
\lambda_1(\kappa H) \geq \kappa^2\;.
\end{align}
Furthermore, with $(\psi, \vec{A})$ being a minimizer of the Ginzburg-Landau functional, we may assume that
\begin{align}
\kappa^2 \| \psi \|_2^2 > Q_{\kappa H \vec{A}}[\psi]\;.
\end{align}
We define 
\begin{align}
\Delta &:= \kappa^2 \| \psi \|_2^2 - Q_{\kappa H \vec{A}}[\psi]\;.
\end{align}
Notice, that 
the GL-equation gives
\begin{align}
\label{eq:4}
 \| \psi\|_4^4 = \frac{\Delta}{\kappa^2}\;.
\end{align}
%

Since $\overline{H_{C_3}(\kappa)} >H > \overline{H_{C_3}^{\rm loc}(\kappa)}$ and using the known asymptotics, Theorem~\ref{thm:GeneralHG3}, we are in a situation where Corollary~\ref{CorNL}, can be applied. Therefore, we get by \eqref{2p} with $p=4$
\begin{align}
\label{eq:BeforeL2}
\| \psi \|_2 \leq C \kappa^{-\frac 14}\| \psi \|_4\;.
\end{align}
Coming back to \eqref{eq:4}, we get
\begin{align}
\label{eq:L2}
\| \psi \|_2 \leq C \kappa^{-\frac{3}{4}} \Delta^{\frac{1}{4}}\;.
\end{align}
We now estimate as in the proof of Theorem~\ref{thm:unifSpecMod} (with the notation $\Phi$, $\vec{a}$ and $b=\curl \vec{A} -1$ as in that proof)
\begin{align}
\label{eq:GaarDet}
0 < \Delta \leq \big[ \kappa^2 - (1-\rho) \lambda_1(\kappa H) \big] \| \psi \|_2^2
+ \rho^{-1} (\kappa H)^2 \int_{\Omega} | \vec{a} \psi |^2 \,dx\;,
\end{align}
for all $0<\rho$.\\
Notice that, by elliptic estimates,
$$
\| \vec{a} \|_{W^{1,2}(\Omega)} \leq \| \Phi \|_{W^{2,2}(\Omega)} \leq C \| b \|_2\;,
$$
so (by the Sobolev estimates in $2$-dimensions), $\|\vec{a}\|_4 \leq C  \| b \|_2$, whereby
\begin{align}
\label{eq:Newa}
(\kappa H)^2 \|\vec{a}\|_4^2 \leq C (\kappa H)^2 \| \curl A -1 \|_2^2 \leq C \Delta\;.
\end{align}
Here we used that ${\mathcal E}_{\kappa,H}[\psi,\vec{A}] \leq 0$ to get the last estimate.

We now insert \eqref{eq:Newa}, \eqref{eq:4}, and \eqref{eq:L2} in \eqref{eq:GaarDet}.
\begin{align}
0 < \Delta &\leq \big[ \kappa^2 - (1-\rho) \lambda_1(\kappa H) \big] \| \psi \|_2^2
+ \rho^{-1} (\kappa H)^2 \| \vec{a} \|_4^2 \|\psi \|_4^2 \nonumber\\
&\leq \big[ \kappa^2 - \lambda_1(\kappa H) \big] \| \psi \|_2^2
+ C \rho \lambda_1(\kappa H) \Delta^{\frac{1}{2}} \kappa^{-\frac{3}{2}}
+ C \rho^{-1} \Delta \frac{\sqrt{\Delta}}{\kappa}
\end{align}
Upon choosing
$\rho = \sqrt{\Delta} \kappa^{-\frac{3}{4}}$, and using that $\lambda_1(\kappa H) < C \kappa^2$, we find
\begin{align}
0 < \Delta &\leq \big[ \kappa^2 - \lambda_1(\kappa H) \big] \| \psi \|_2^2
+ C \Delta \kappa^{-\frac{1}{4}}\;.
\end{align}
 When $\kappa$ is so big that $C \kappa^{-\frac{1}{4}} <1$, we therefore get
\begin{align}
0 < (1- C \kappa^{-\frac{1}{4}}) \Delta 
\leq \big[ \kappa^2 - \lambda_1(\kappa H) \big] \| \psi \|_2^2\;.
\end{align}
But this is in contradiction with the initial hypothesis \eqref{eq:Contradiction}.
This finishes the proof of Theorem \ref{thm:Identical}.\\
\qed

\appendix
\section{General non-degenerate domains}
\label{AppGenNonDeg}

Assumption~\ref{assump:omega} is unnecessarily restrictive for some of our results. Consider the alternative setup given by Assumption~\ref{assump:GenNondegen}.
In this appendix we will prove an asymptotics for the low lying eigenvalues of the magnetic Neumann operator on a domain $\Omega$ satisfying Assumption~\ref{assump:GenNondegen}.
For convenience of comparison with the work \cite{FournaisHelffer2} we consider the semi-classical asymptotics, i.e. we consider the operator
\begin{align}
\label{eq:Kh}
{\mathcal K}_h = (-ih \nabla - \vec{A})^2\;,
\end{align}
with Neumann boundary conditions, and will study the asymptotics of the spectrum\footnote{Clearly, as in Subsection~\ref{CalcAsymp}, this is equivalent by scaling to the large $B$ asymptotics of $\Spec {\mathcal H}(B)$.} $\{ \mu^{(n)}(h)\}_{n=1}^{\infty}$ of ${\mathcal K}_h $ as $h \rightarrow 0_{+}$. Here $\vec{A}$ is a vector potential generating a unit magnetic field, i.e. $\curl \vec{A} =1$. It will be practical to have a globally defined choice of $\vec{A}$, so we will work with the special choice $\vec{A} = \frac{1}{2}(-x_2, x_1)$.

Let $\delta = \frac{1}{4} \min_{j \neq k} |s_j - s_k|$ and let $\gamma : {\mathbb R}/|\partial \Omega| \rightarrow \partial \Omega$ be the standard parametrization of $\partial \Omega$.
For $j \in \{1, \ldots, N\}$, let $\Omega^{(j)}$ be a bounded domain with smooth boundary satisfying that
\begin{enumerate}
\item \label{et}There exists a (smooth) parametrization $\gamma^{(j)}$ of $\partial \Omega^{(j)}$ such that 
$\gamma^{(j)}(s) = \gamma(s-s_j)$ for $s \in (-\delta, \delta)$.
\item \label{to} $\gamma^{(j)}(0)$ is the unique point of maximum curvature of $\partial \Omega^{(j)}$.
\end{enumerate}
In particular, the domains $\Omega^{(j)}$ satisfy the Assumption~\ref{assump:omega}.

Define ${\mathcal K}_h^{(j)}$ to be the differential operator from \eqref{eq:Kh} defined on $\Omega^{(j)}$ with Neumann boundary conditions, and let $\{ \mu^{(n,j)}(h)\}_{n=1}^{\infty}$ be the eigenvalues (in non-decreasing order) of ${\mathcal K}_h^{(j)}$. From \cite{FournaisHelffer2} we get the following description of the $\mu^{(n,j)}(h)$'s.

\begin{thm}
\label{thm:AsymptoticsGenNondegen}~\\
Suppose $\Omega$ satisfies Assumption~\ref{assump:GenNondegen}.
Let $j \in \{1, \ldots, N\}$.
For all $n \in {\mathbb N}\setminus \{0\}$, there exists 
a sequence $\{ \zeta_{\ell}^{(n,j)}\}_{\ell=1}^{\infty} \subset {\mathbb R}$ (which can be calculated recursively to any order) such that $\mu^{(n,j)}(h)$ admits the following asymptotic expansion (for $h \searrow 0$)~:
\begin{align}
\mu^{(n,j)}(h)\sim \Theta_0 h - k_{\rm max} C_1 h^{\frac{3}{2}} + C_1
\Theta_0^{\frac{1}{4}} \sqrt{\tfrac{3 k_{2,j}}{2}} (2n-1)h^{\frac{7}{4}}  + h^{\frac{15}{8}}\sum_{\ell=0}^\infty  h^{\frac{\ell}{8}} \zeta_{\ell}^{(n,j)}\;.
\end{align}
Furthermore, the coefficients $\{ \zeta_{\ell}^{(n,j)}\}_{\ell=1}^{\infty}$ are independent of the choice of $\Omega^{(j)}$ satisfying (\ref{et}) and (\ref{to}) above.
\end{thm}

Let now $\{ \tilde{\mu}^{(n)}(h)\}_{n=1}^{\infty}$ be the sequence of the $\{ \mu^{(n,j)}(h) \}$, $j=1, \ldots, N$, $n=1,\ldots, \infty$ with multiplicity and in non-decreasing order. In this context, it is convenient to consider the
operator ${\mathcal K}_h^{\rm comb}$ defined as ${\mathcal K}_h^{(1)} \oplus \cdots \oplus {\mathcal K}_h^{(N)}$ as an operator on $L^2(\Omega^{(1)}) \oplus \cdots\oplus L^2(\Omega^{(N)})$. Then clearly $\tilde{\mu}^{(n)}(h)$ is simply the $n$'th eigenvalue of ${\mathcal K}_h^{\rm comb}$.

The main result of this appendix is the following theorem.

\begin{thm}
\label{thm:genNondegen}~\\
Suppose that $\Omega$ satisfies Assumption~\ref{assump:GenNondegen}.
With the above notation, we have for all $n \in {\mathbb N} \setminus \{ 0 \}$,
$$
\mu^{(n)}(h) - \tilde{\mu}^{(n)}(h)  = {\mathcal O}(h^{\infty})\;.
$$
\end{thm}

By simple manipulations we convert the small $h$ asymptotics to a large $B$ asymptotics.

\begin{cor}
\label{cor:LargeB}~\\
Suppose $\Omega$ satisfies Assumption~\ref{assump:GenNondegen} and let $\lambda_1(B)$ be the smallest eigenvalue of ${\mathcal H}(B)$. Then there exists a sequence $\{\zeta_j\}_{j=0}^{\infty} \subset {\mathbb R}$ such that for all $M>0$,
\begin{align}
\label{eq:BlargeGen}
\lambda_1(B) = \Theta_0 B - C_1 k_{\rm max} B^{\frac{1}{2}} + C_1 \sqrt[4]{\Theta_0}\sqrt{\tfrac{3k_2}{2}}B^{\frac{1}{4}} +
B^{\frac{1}{8}} \sum_{j=0}^M \zeta_j B^{-\frac{j}{8}} + {\mathcal O}(B^{-\frac{M}{8}}).
\end{align}
\end{cor}

\begin{proof}[Proof of Theorem~\ref{thm:genNondegen}]~\\
We may choose $\eta >0$ such that
\begin{align}
&B(\gamma(s_j), 2\eta) \cap B(\gamma(s_k), 2\eta) = \emptyset \text{ for } j \neq k\;,\nonumber\\
&B(\gamma(s_j), 2\eta) \cap \Omega = B(\gamma(s_j), 2\eta) \cap \Omega^{(j)}\;.
\end{align}
Let $\phi_j$ be a smooth function satisfying
\begin{align}
\phi_j(x) &= 1 \text{ on } B(\gamma(s_j), \eta)\;,&
\supp \phi_j & \subset B(\gamma(s_j), 2\eta)\;.
\end{align}
Let $\psi^{(n)}_h$ be the $n$'th eigenfunction of ${\mathcal K}_h$. Furthermore, let $\psi^{(n,j_n)}_h$ be the eigenfunction of ${\mathcal K}^{(j_n)}_h$ corresponding to $\tilde{\mu}^{(n)}(h)$, here $j_n$ may depend on $h$. 

We may consider $\phi_j \psi^{(n,j_n)}_h$ as a function on $\Omega$ (extended by zero). By the Agmon estimates (in the $\Omega^{(j)}$'s), we easily get
\begin{align}
&\langle \phi_{j_n} \psi^{(n,j_n)}_h \, | \,  \phi_{j_m} \psi^{(m,j_{m})}_h \rangle = \delta_{m,n}+ {\mathcal O}(h^{\infty})\;, \nonumber\\
&\langle \phi_{j_n} \psi^{(n,j_n)}_h \, | \, {\mathcal K}_h \phi_{j_m} \psi^{(m,j_{m})}_h \rangle = \delta_{m,n} \tilde{\mu}^{(n)}(h) + {\mathcal O}(h^{\infty})\;.
\end{align}
Therefore, the variational characterization of eigenvalues gives
\begin{align}
\mu^{(n)}(h) \leq \tilde{\mu}^{(n)}(h) + {\mathcal O}(h^{\infty})\;.
\end{align}

We now prove the opposite inequality. Define $\phi = \sum_{j=1}^N \phi_j$. For $\psi \in L^2(\Omega)$, we can naturally identify $\phi \psi$ with an element of $L^2(\Omega^{(1)}) \oplus \cdots\oplus L^2(\Omega^{(N)})$. We will do so without changing the notation. 
Using again the Agmon estimates (this time in $\Omega$), we see that
\begin{align}
&\langle \phi \psi^{(n)}_h \, | \,  \phi \psi^{(m)}_h \rangle = \delta_{m,n}+ {\mathcal O}(h^{\infty})\;, \nonumber\\
&\langle \phi \psi^{(n)}_h \, | \, {\mathcal K}_h^{\rm comb} \phi \psi^{(m)}_h \rangle = \delta_{m,n} \tilde{\mu}^{(n)}(h) + {\mathcal O}(h^{\infty})\;.
\end{align}
Here the inner products on the left hand sides are the natural inner products on $L^2(\Omega^{(1)}) \oplus \cdots\oplus L^2(\Omega^{(N)})$.

A second application of the variational principle therefore gives 
\begin{align}
\tilde{\mu}^{(n)}(h) \leq \mu^{(n)}(h) + {\mathcal O}(h^{\infty})\;,
\end{align}
and finishes the proof.
\end{proof}

\section{Boundary coordinates}
\subsection{The coordinates}\label{bdry}~\\ 
Let $\Omega$ be a smooth, simply-connected domain in ${\mathbb R}^2$.
Let $\gamma: {\mathbb R}/|\partial \Omega| \rightarrow \partial \Omega$ be a parametrization of the boundary with $|\gamma'(s)|=1$ for all $s$. Let $\nu(s)$ be the unit vector, normal to the boundary, pointing inward at the point $\gamma(s)$. We choose the orientation of the parametrization $\gamma$ to be counter-clockwise, so
\begin{align*}
\det\big( \gamma'(s), \nu(s)\big) = 1\;.
\end{align*}
The curvature $k(s)$ of $\partial \Omega$ at the point $\gamma(s)$ is now defined by
\begin{align*}
\gamma''(s) = k(s) \nu(s)\;.
\end{align*}
The map $\Phi$ defined in the introduction, 
\begin{align}
\label{eq:t0}
&\Phi : {\mathbb R}/|\partial \Omega| \times (0,t_0) \rightarrow \Omega\;, \\
&(s,t) \mapsto  \gamma(s) + t \nu(s)\;,
\end{align}
is, when $t_0$ is sufficiently small, clearly a diffeomorphism with image
$$
\Phi\big( {\mathbb R}/|\partial \Omega| \times (0,t_0) \big) = 
\{ x \in \Omega \big | \dist(x, \partial \Omega) < t_0 \} =: \Omega_{t_0}\;.
$$
Furthermore, $t(\Phi(s,t)) = t$.

If $\vec{A}$ is a vector field on $\Omega_{t_0}$ with $B = \curl \vec{A}$ we define the associated fields in $(s,t)$-coordinates by
\begin{align}
\label{eq:tilde}
\tilde{A}_1(s,t) &= (1-tk(s)) \vec{A}(\Phi(s,t)) \cdot \gamma'(s)\;, &
\tilde{A}_2(s,t) &= \vec{A}(\Phi(s,t)) \cdot \nu'(s)\;, \\
\tilde{B}(s,t) &= B(\Phi(s,t))\;.&&
\end{align}
 Then $\partial_s \tilde{A}_2 - \partial_t \tilde{A}_1 = (1-tk(s)) \tilde{B}$. Furthermore, for all $u \in W^{1,2}(\Omega_{t_0})$, we have, with $v = u \circ \Phi$,
\begin{align}
\label{eq:Quad}
&\int_{\Omega_{t_0}} |(-i \nabla - \vec{A}) u|^2\,dx \\
&\quad\quad=
\int \big\{ (1-tk(s))^{-2} \Big|(-i\partial_s - \tilde{A}_1) v\big|^2 + \big|(-i\partial_t - \tilde{A}_2) v\big|^2\Big\} (1-tk(s)) \,dsdt\;, \nonumber\\
&\int_{\Omega_{t_0}} | u(x)|^2\,dx = \int |v(s,t)|^2 (1-tk(s))\,dsdt\;.\nonumber
\end{align}

\begin{lemma}
\label{lem:GoodGaugeImp}~\\
Suppose $\Omega$ is a bounded, simply connected domain with smooth boundary and let $t_0$ be the constant from \eqref{eq:t0}. Then there exists a constant $C>0$ such that, if 
$\vec{A}$ is a vector potential in $\Omega$ with
\begin{align}
\curl \vec{A} =1 \quad \text{ on } \partial \Omega\;,
\end{align}
and with $\tilde{A}$ defined as in \eqref{eq:tilde}, then there exists a gauge function $\varphi(s,t)$ on ${\mathbb R}/|\partial \Omega| \times (0,t_0)$ such that
\begin{align}
\label{eq:Top}
\bar{A}(s,t) = \begin{pmatrix} \bar{A}_1(s,t) \\ \bar{A}_2(s,t) \end{pmatrix} :=
\tilde{A}-\nabla_{(s,t)} \varphi =
\begin{pmatrix} \gamma_0 - t + \frac{t^2k(s)}{2} + t^2 b(s,t) \\ 0 \end{pmatrix}\;,
\end{align}
where 
\begin{align*}
\gamma_0 = \frac{1}{|\partial \Omega|}\int_{\Omega} \curl \vec{A}\,dx\;,
\end{align*}
and $b$ satisfies the estimate,
\begin{align*}
\| b \|_{L^{\infty}({\mathbb R}/|\partial \Omega| \times (0,\frac{t_0}{2}))} &\leq C \| \curl \vec{A}-1\|_{C^{1}(\Omega_{t_0})} \;.
\end{align*}
Furthermore, if $[s_0, s_1]$ is a subset of ${\mathbb R}/|\partial \Omega|$ with $s_1-s_0 < |\partial \Omega|$, then we may choose $\varphi$ on $(s_0, s_1) \times (0,t_0)$ such that
\begin{align}
\label{eq:VeryGoodGauge}
\bar{A}(s,t) = \begin{pmatrix} \bar{A}_1(s,t) \\ \bar{A}_2(s,t) \end{pmatrix} :=
\tilde{A}-\nabla_{(s,t)} \varphi =
\begin{pmatrix}  - t + \frac{t^2k(s)}{2} + t^2 b(s,t) \\ 0 \end{pmatrix}\;.
\end{align}
\end{lemma}

\begin{proof}~\\
Notice first that
\begin{align*}
\int_0^{|\partial \Omega|} A_1(s,0)\,ds = \int_0^{|\partial \Omega|} \vec{A} \cdot \gamma'(s)\,ds
= \int_{\Omega} \curl \vec{A}\,dx\;.
\end{align*}
Let us write 
\begin{align*}
\nu &= \curl \vec{A} -1\,,&
\tilde{\nu}(s,t) &= \nu(\Phi(s,t))\;,&
\tilde{\nu}' = \frac{\tilde{\nu}}{t}\;.
\end{align*}
Then $\| \tilde{\nu}' \|_{L^{\infty}} \leq C \| \nu\|_{C^{1}(\Omega_{t_0})}$ and
\begin{align*}
\partial_s \tilde{A}_2 - \partial_t \tilde{A}_1 = (1-tk(s))(1+t\tilde{\nu}')\;.
\end{align*}
Define 
\begin{align}
\label{eq:varphi}
\varphi(s,t) = \int_0^t \tilde{A}_2(s,t')\;dt' + \big( \int_0^s \tilde{A}_1(s',0)\,ds' - s \gamma_0\big)\;.
\end{align}
Then $\varphi$ is a well-defined continuous function on ${\mathbb R}/|\partial \Omega| \times (0,t_0)$. We pose $\bar{A} = \tilde{A} - \nabla \varphi$ and find
\begin{align*}
&\bar{A}(s,t) = \begin{pmatrix} \bar{A}_1(s,t) \\ \bar{A}_2(s,t) \end{pmatrix}  = 
\begin{pmatrix} \bar{A}_1(s,t) \\ 0 \end{pmatrix}\;, \\
&\partial_t \bar{A}_1(s,t) = -(\partial_s \tilde{A}_2 - \partial_t \tilde{A}_1) = -(1-tk(s))(1+t\tilde{\nu}')\;,\\
&
\bar{A}_1(s,0)= \gamma_0\;.
\end{align*}
Therefore,
\begin{align*}
\bar{A}_1(s,t) = \gamma_0 - t + \frac{t^2 k(s)}{2} - \int_0^t t' (1-t'k(s)) \tilde{\nu}'(s,t')\;dt'\;,
\end{align*}
and we get \eqref{eq:Top} by applying l'H\^{o}pital's rule to the integral.

In the case where we only consider a simply connected part $(s_0, s_1) \times (0,t_0)$ of the ring ${\mathbb R}/|\partial \Omega| \times (0,t_0)$, we have trivial topology and therefore any two vector fields generating the same magnetic field are gauge equivalent. Therefore the constant term, $\gamma_0$, can be omitted. 
From a more practical point of view, one can see that we can omit the term $s\gamma_0$ in \eqref{eq:varphi} since we do not need to ensure periodicity of the function $\varphi$.\\
\end{proof}

\subsection{The model operator}~\\
\label{AppA}
When considering functions localized near the boundary (i.e. $t$ small), and after making a partial Fourier transformation in the $s$-variable, one is led from the quadratic form in \eqref{eq:Quad} to the study of a simpler operator on the half-line depending on a real parameter $\zeta$
\begin{align}
h(\zeta) := -\frac{d^2}{d\tau^2} + (\zeta + \tau)^2\;,
\end{align}
on $L^2({\mathbb R}_{+},d\tau)$. The boundary condition at $\tau=0$ is the usual Neumann boundary condition.
It is clear that this (self adjoint) operator is very important for the subject considered in the present paper, and it has been extensively studied.
We will here recall the main spectral properties (see \cite{DaHe} and \cite{BeSt}) of $h(\zeta)$. 
We denote by $\mu(\zeta)$ the lowest eigenvalue of $h(\zeta)$ and by $\varphi_{\zeta}$ the
corresponding strictly positive normalized eigenfunction. Then one has the following statements.

\begin{itemize}
\item The infimum, 
$\inf_{\zeta\in {\mathbb R}} \inf \Spec ( h(\zeta)$, is actually a minimum: there exists
$\xi_0 < 0$ such that  $ \mu ( \xi )$ decreases monotonically to a minimum value 
$ \frac{1}{2} < \Theta_0 <1$ and then increases monotonically again.
\item $\Theta_0 = \xi_0^2$. 
\end{itemize}
We will write $u_0$ instead of $\phi_{\xi_0}$ and define
\begin{align}\label{eq:C1}
C_1 = \frac{u_0^2(0)}{3}\;.
\end{align}

%

\end{document}